\documentclass[twocolumn,
superscriptaddress,
aps,
prb,
]{revtex4-2}

\usepackage{graphicx}% Include figure files
\usepackage{dcolumn}% Align table columns on decimal point
\usepackage{bm}% bold math
\usepackage[colorlinks=true,urlcolor=blue,citecolor=blue]{hyperref}

\usepackage{amsmath}
\usepackage{amssymb}
\usepackage{physics}
\usepackage{ulem}

\begin{document}

%\preprint{APS/123-QED}

\title{Phases and Phase Transitions of the Disordered Quantum Clock Model}% Force line breaks with \\

\author{Pulloor Kuttanikkad Vishnu}
 \affiliation{Department of Physics, Indian Institute of Technology Madras, Chennai 600036, India.}%

\author{Gaurav Khairnar}%
 \affiliation{Department of Physics, Missouri University of Science and Technology, Rolla, Missouri 65409, USA.}%

\author{Rajesh Narayanan}
  \email{rnarayanan@zmail.iitm.ac.in}
 \affiliation{Department of Physics, Indian Institute of Technology Madras, Chennai 600036, India.}%

\author{Thomas Vojta}
  \email{vojtat@mst.edu}
 \affiliation{%
 Department of Physics, Missouri University of Science and Technology, Rolla, Missouri 65409, USA.}%

\date{\today}% It is always \today, today, but any date may be explicitly specified

\begin{abstract}
We study the phases and phase transitions of a disordered one-dimensional quantum $q$-state clock Hamiltonian using large-scale Monte Carlo simulations. Making contact with earlier results, we confirm that the clean, translational invariant version of the model, for $q=6$, hosts an intermediate emergent quasi-long-range ordered (QLRO) phase between the symmetry-broken true long-range ordered (TLRO) phase and the disordered (paramagnetic) phase. With increasing disorder strength, the quasi-long-range ordered phase shrinks and finally vanishes at a multi-critical point, beyond which there is a direct transition from the TLRO phase to the paramagnetic phase. After establishing the phase diagram, we characterize the critical behaviors of the various quantum phase transitions in the model. We find that weak disorder is an irrelevant perturbation of the Berezinskii-Kosterlitz-Thouless transitions that separate the QLRO phase from the TLRO and paramagnetic phases. For stronger disorder, some of the critical exponents become disorder-dependent already before the system reaches the multicritical point. We also show that beyond the multicritical point, the direct transition from the TLRO phase to the paramagnetic phase is governed by an infinite-randomness critical point in line with strong-disorder renormalization group predictions. While our numerical results are for $q=6$, we expect the qualitative features of the behavior to hold for all $q>4$.
\end{abstract}

\maketitle

%\tableofcontents
%%%%%%%%%%%%%%%%%%%%%%%%%%%%%%%%%%%%%%%%%%%%%%%%%%%%%%%%%%%%%%%%%%%%%%%%%%%%%%%%%%%
\section{Introduction}
\label{sec-1}
%%%%%%%%%%%%%%%%%%%%%%%%%%%%%%%%%%%%%%%%%%%%%%%%%%%%%%%%%%%%%%%%%%%%%%%%%%%%%%%%%%%

The impact of quenched disorder on both quantum and classical phase transitions is a subject matter with a long and rich history (see, e.g., Ref.\ \cite{vojta_jpamg_06, *Vojta13, *vojta_arcmp_19} for recent reviews). These studies have produced a welter of results ranging from the celebrated Harris \cite{harris_jpcssp_74} and Imry-Ma
\cite{PhysRevLett.35.1399} criteria that predict the stability of phases and phase transitions against quenched impurities to the discoveries of numerous exotic phenomena close to disordered quantum phase transitions. These include infinite-randomness critical points and the attendant quantum Griffiths phases \cite{fisher_prl_92,*fisher_prl_95,RiegerYoung96,GuoBhattHuse96, hoyos_kotabage_prl_07,*vojta_kotabage_prb_09} as well as smeared phase transitions \cite{Vojta03a, *vojta_jpamg_03, *HoyosVojta08}.

The vast majority of work within this area has centered on studying the effects of impurity-induced randomness on phases that follow the Landau symmetry classification and the corresponding symmetry-breaking phase transitions. Relatively little is known about the effects of random disorder on other types of transitions including topological
transitions that do not break the symmetry of a local order parameter. A well-known example is the Berezinskii-Kosterlitz-Thouless (BKT) transition \cite{Berezinskii71,kosterlitz_thouless_jpcssp_73} that
frequently obtains in (1+1)-dimensional quantum many-body systems with $O(2)$ [or $U(1)$] order parameter symmetry or, equivalently, in the corresponding classical statistical models in two space dimensions, such as the two-dimensional XY model. These transitions correspond to the binding and unbinding of vortex/anti-vortex pairs and demarcate a phase with exponentially decaying order parameter correlations from a critical, quasi-long-range ordered (QLRO) phase characterized by power-law decay of the correlations.

\begin{figure}
    \centering
    \includegraphics[width=\columnwidth]{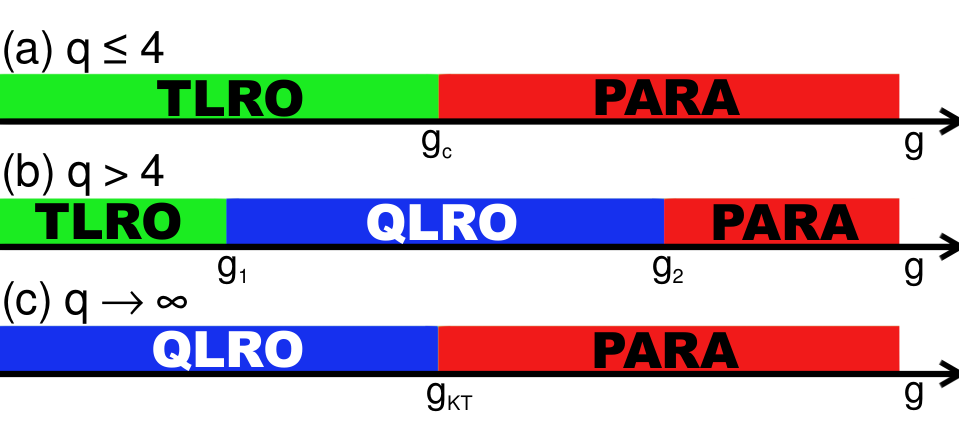}
    \caption{Phases of the one-dimensional $q$-state quantum clock model for (a) $q\le4$ (b) $q>4$ and (c) $q\rightarrow\infty$ in the absence of quenched randomness. $g$ denotes the parameter used to tune the system through the transition.}
    \label{clean-quantum-phases}
\end{figure}

Interestingly, such critical or massless phases, accompanied by BKT transitions, can also occur when the $O(2)$ order parameter symmetry is broken down
into $q$ discrete values, evenly spaced on the unit circle, resulting in a discrete $Z_q$ symmetry. Extensive studies of the ferromagnetic $q$-state quantum clock model in one space dimension (and its two-dimensional classical analog) \cite{jose_kadanoff_prb_78, elitzur_pearson_prd_79, cardy_jpamg_80, tobochnik_prb_82, challa_landau_prb_86, yamagata_ono_jpamg_91, tomita_okabe_prb_02, lappili_pfeifer_prl_06,matsuo_nomura_jpamg_06, lappili_pfeifer_prl_06, brito_redinz_pre_10, enting_clisby_jsp_11,borisenko_cortese_pre_11, ortiz_cobanera_npb_12, kumano_hukushima_prb_13, chatelain_jsmte_14,  chen_lia_cpsipl_17,sun_vekua_prb_19,li_yang_pre_20,miyajima_murata_prb_21, chen_hou_pre_22,tuan_nui_pre_22,LiPaiGu22,otsuka_shiina_jpamt_23,guo_he_njp_23},  have shown that such systems host a QLRO phase that intervenes between a symmetry-broken truly long-range ordered (TLRO) phase, and a disordered (paramagnetic, in the language of magnetic systems) phase for all $q>4$. This is illustrated in  Fig.~\ref{clean-quantum-phases}.
In contrast, a direct transition from the symmetry-broken to the disordered phase obtains for all $q\le4$.

Physical realizations of the $q$-state clock model are manifold: For instance, the planar-to-buckled instability in two-dimensional lattices of ions can be understood via a mapping to a six-state clock model \cite{podolsky_shimshoni_prx_16}. Other physical manifestations of the six-state clock model include the half-filled extended Hubbard model on a triangular lattice in the atomic limit \cite{kaneko_nonomura_prb_18}, the displacive structural phase transition in certain two-dimensional solids \cite{vink_pre_18}, the two-dimensional frustrated Heisenberg antiferromagnet on a windmill lattice \cite{orth_chandra_prl_12,orth_chandra_prb_14, bilahari_chandra_prl_15}, the Blume-Capel antiferromagnet on a triangular lattice \cite{zukovic_bobal_pre_13}, a triangular-lattice antiferromagnetic Ising model with a spatially anisotropic next-nearest-neighbor ferromagnetic coupling \cite{otsuka_okabe_pre_06}, stacked triangular antiferromagnetic Ising models \cite{blankschtein_ma_prb_84,isakov_moessner_prb_03}, BKT phase transitions in a Kagome spin ice system \cite{chern_techernyshyov_ptrsa_12,su_hu_prb_23}, the melting of three-sublattice order in triangular and Kagome antiferromagnets \cite{damle_prl_15}, the domain pattern in certain layered hexagonal materials \cite{chae_lee_prl_12}, and the melting of magnetic order in frustrated triangular rare earth magnets \cite{li_liao_nc_20}.

The impact of (weak) quenched disorder on a system exhibiting a BKT transition was studied in the seminal work by Giamarchi and Schulz \cite{giamarchi_schulz_el_87}, harnessing a perturbative renormalization group (RG). This approach was later extended to second order in the disorder strength by Ristivojevic et al.\ \cite{ristivojevic_petkovic_prl_12}. Set in the context of the superfluid-insulator transition of disordered 1D bosons, these studies concluded that the BKT transition features universal critical exponents and a universal value of the Luttinger parameter. In contrast, a strong-disorder renormalization group (SDRG) calculation by Altman et al.\ \cite{AKPR04,AKPR08,altman_kafri_prb_10} predicted that the superfluid-insulator transition belongs to a disordered BKT universality class featuring non-universal, disorder-dependent exponents and a non-universal value of the Luttinger parameter, at least for sufficiently strong quenched disorder. In the following years, the question of whether or not quenched disorder can change the critical behavior of a BKT transition and, if so, the nature of the resulting transition was controversially discussed in the literature (see, e.g.\, Refs.\ \cite{PielawaAltman13,PolletProkofevSvistunov14,Doggenetal17,PfefferYaoPollet19,LemarieMaccariCastellani19} and references therein). To the best of our knowledge, a full resolution to this problem has not been achieved, yet. However, there seems to be a consensus that sufficiently strong disorder can lead to a transition that is different from the Giamarchi-Schulz scenario and governed by the physics of broadly distributed weak links. Most of the above work considered the one-dimensional Bose gas problem with diagonal disorder. Hrahsheh and Vojta \cite{hrahsheh_vojta_prl_12} performed large-scale Monte-Carlo simulations for the particle-hole symmetric case with the off-diagonal disorder (which maps onto a classical XY-model with the columnar disorder). In analogy with the Bose glass results, they found that the critical behavior is universal (and identical to the clean case) for the weak disorder but becomes non-universal and disorder-dependent for the stronger disorder.

\begin{figure}
    \centering
    \includegraphics[width=\columnwidth]{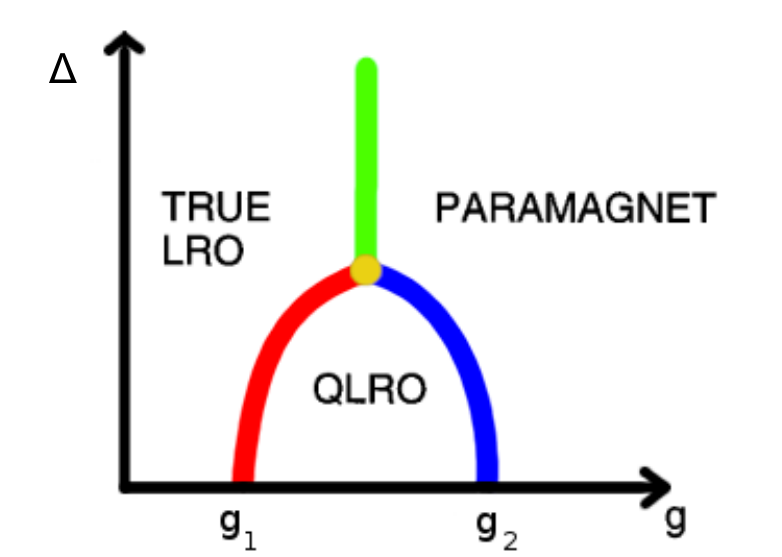}
 \caption{Schematic phase diagram for disordered quantum clock model as proposed in Ref.\ \cite{senthil_majumdar_prl_96}. $g$ denotes the quantum tuning parameter, and $\Delta$ is the disorder strength.}
 \label{fig:Senthil-phasediagram}
\end{figure}

The effect of quenched impurities on the one-dimensional quantum clock model was analyzed by Senthil and Majumdar by laying recourse to an SDRG treatment \cite{senthil_majumdar_prl_96}. Their results, valid for sufficiently strong randomness, yield a direct transition of infinite-randomness type from the disordered (paramagnetic) phase to the symmetry-broken (TLRO) phase. This suggests that the size of the intermediate QLRO phase decreases with increasing disorder strength, leading to the schematic phase diagram in Fig.\  \ref{fig:Senthil-phasediagram}, proposed in Ref.\  \cite{senthil_majumdar_prl_96}.
At lower disorder strengths (not compatible with the SDRG technique), an earlier DMRG study showed the existence of disorder-dependent exponents across the transition from the QLRO phase to the paramagnetic phase \cite{CarlonLajkoIgloi01}.

In the present manuscript, we report the results of a comprehensive study of the one-dimensional disordered quantum clock model (focusing on $q=6$ as a prototypical case for all $q>4$) by means of Monte Carlo simulations. We construct the phase diagram as a function of quantum fluctuation strength and disorder, spanning the range from the clean limit all the way to the strong disorder regime. The paper is organized as follows: In Sec.\ \ref{sec-2}, we introduce the quantum clock Hamiltonian and its mapping to a two-dimensional classical clock model. Details of the Monte-Carlo simulations are presented in Sec.\ \ref{sec-3} along with a discussion of the various observables and their expected scaling behavior.
In Sec.\ \ref{sec-4}, we discuss the results of our simulations, which encompasses a discussion of the phase diagram as well the properties of the various phase transitions in the weak and strong-disorder regimes. Finally, we set forth our conclusions in Sec.\ \ref{sec-5}.

%%%%%%%%%%%%%%%%%%%%%%%%%%%%%%%%%%%%%%%%%%%%%%%%%%%%%%%%%%%%%%%%
\section{Model}
\label{sec-2}
%%%%%%%%%%%%%%%%%%%%%%%%%%%%%%%%%%%%%%%%%%%%%%%%%%%%%%%%%%%%%%%%%

To define the Hamiltonian of the one-dimensional disordered quantum clock model, consider a chain of lattice sites $j$.
Each site is occupied by a $q$-state clock variable having eigenstates $\ket{p_j}$ with $p_j =0, 1, \ldots, q-1$. Written in
the clock-state basis, the Hamiltonian reads
\begin{eqnarray}
      H=-\sum_{j}J_{j}\cos\left[\frac{2\pi ({p}_{j}-{p}_{j+1})}{q}\right]
      -\sum_{j}\frac{h_{j}}{2} \left(\hat{\Gamma}_{j}+\hat{\Gamma}^{\dagger}_{j}\right)~.
      \label{eqn:quantum-hamiltonian}
\end{eqnarray}
The raising and lowering operators are defined by $\hat{\Gamma}^\dagger\ket{p}=\ket{(p+1)\mod q}$ and $\hat{\Gamma}\ket{p}=\ket{(p-1)\mod q}$.
The energies $J_{j}$ and $h_{j}$ represent the nearest-neighbor couplings and ``transverse field'' terms, respectively, at site $j$. In the limit $q \rightarrow \infty$, the Hamiltonian maps onto a quantum rotor model whereas the $q = 2$ case reduces to the transverse field Ising model.

In preparation for the Monte Carlo simulations we now employ the quantum-to-classical mapping method (see, e.g.,  Ref.~\cite{Sachdev_book99}) that recasts the partition function of a $d$-dimensional quantum model as the partition function of an equivalent classical model in $D=d+1$ dimensions. In our case, the mapping, which is detailed in Appendix A leads from the one-dimensional quantum Hamiltonian (\ref{eqn:quantum-hamiltonian}) to a classical clock model in $D=2$  space dimensions. The resulting classical Hamiltonian reads

\begin{equation}
\label{eq_Hamiltonian_cl}
\begin{split}
    H_{\rm cl}=&-\sum_{j,\tau} J^{s}_{j}\cos\left[\frac{2\pi(p_{j,\tau}-p_{j+1,\tau})}{q}\right]\\
    &-\sum_{j,\tau}J^{t}_{j}\cos\left[\frac{2\pi(p_{j,\tau}-p_{j,\tau+1})}{q}\right]~.
\end{split}
\end{equation}
Here, the coordinates $j$ and $\tau$ correspond to the real-space and imaginary-time positions, respectively, in the original quantum problem. In the following, we treat interactions $J_j^{s}$ and $J_j^{t}$ as constants and tune the strength of the fluctuations by varying the ``classical'' temperature $T$ of the model (\ref{eq_Hamiltonian_cl}). Thus, the classical temperature plays the role of the quantum tuning parameter $g$ in Fig.\ \ref{fig:Senthil-phasediagram}, and $T_{c1}$ and $T_{c2}$ will denote the transition temperatures from the TLRO phase to the QLRO phase and from the QLRO phase to the disordered (paramagnetic) phase, respectively.

 The universal features of the phase diagram and phase transitions are shared  by the quantum model (\ref{eqn:quantum-hamiltonian}) and its classical counterpart (\ref{eq_Hamiltonian_cl}).  Analyzing the classical model by means of classical Monte Carlo simulations allows us to employ cluster algorithms that are highly efficient
 even in the presence of disorder, giving us the ability to study large system sizes that reduce finite-size effects and probe the thermodynamic limit
 \footnote{in contrast, the presence of quenched disorder  tends to curtail the efficiency of other simulation techniques for quantum many-particle systems such as DMRG or tensor-network based methods.}.

\begin{figure}%[H]
    \centering
    \includegraphics[width=\columnwidth]{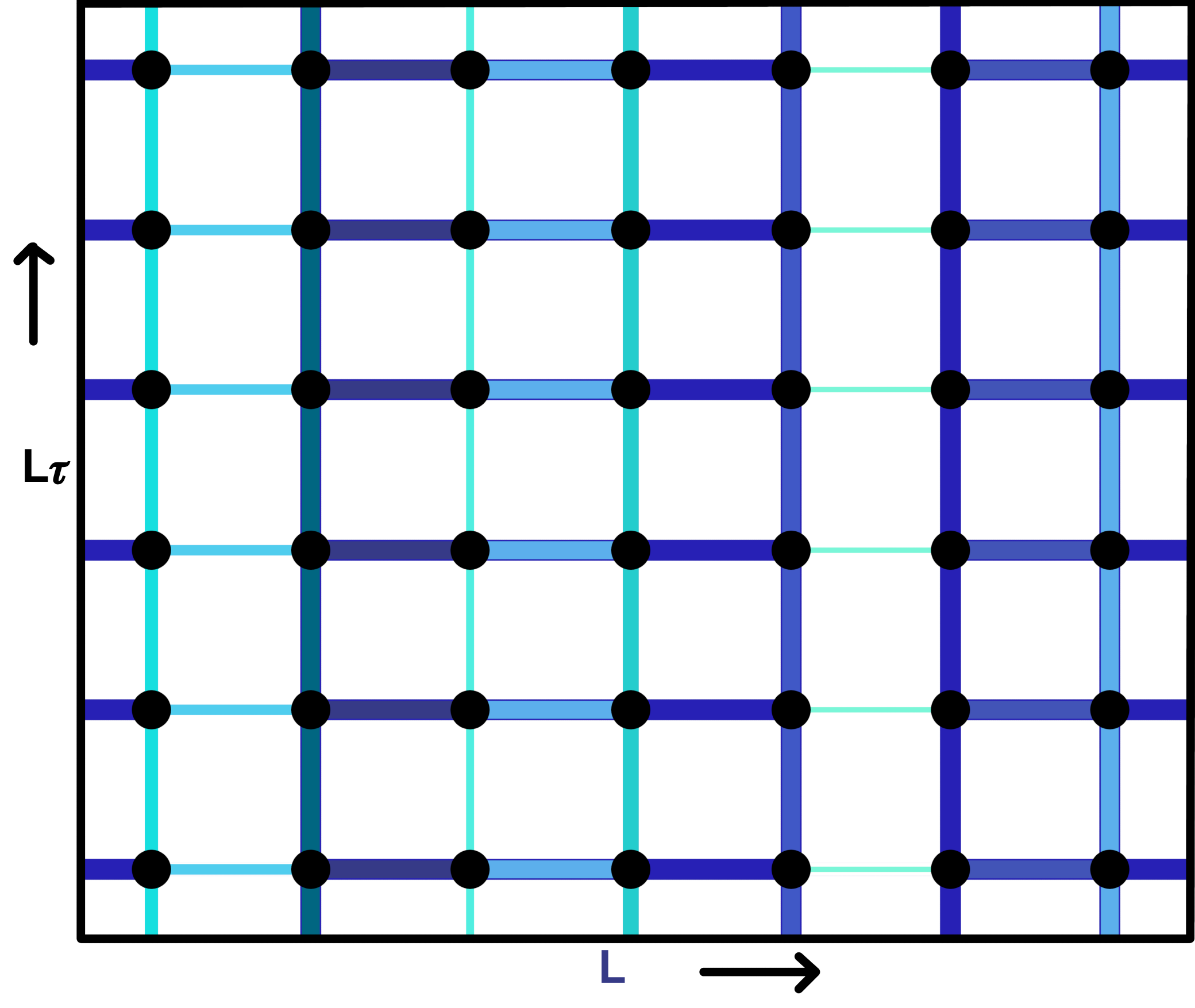}
    \caption{Schematic of a lattice with the correlated disorder in both horizontal (space-like) and vertical (imaginary-time-like) bonds. Different colors (shading) indicate different coupling strengths.}
    \label{fig:lattice-picture}
\end{figure}

Quenched random disorder can be introduced into the quantum Hamiltonian (\ref{eqn:quantum-hamiltonian}) by making
$J_j$ and $h_j$ independent random variables.
The quantum-to-classical mapping implies that such spatially uncorrelated disorder in the quantum Hamiltonian corresponds to columnar disorder in the classical model (\ref{eq_Hamiltonian_cl}), i.e., disorder that is perfectly correlated along the imaginary time direction \cite{mccoy_wu_68,shankar_murty_87}\footnote{For the sake of completeness we refer the reader to \cite{surungan2005kosterlitz}, where the two-dimensional classical clock model is studied in the presence of completely uncorrelated quenched randomness.}. Consequently, both interactions in the classical model,  $J^{s}_{j}$ and $J^{t}_{j}$, are only functions of the spatial index $j$.  A pictorial representation of such a lattice with correlated impurities is shown in Fig.\ \ref{fig:lattice-picture}.
As the disorder breaks the symmetry between the two directions, we need to specify two different linear system sizes $L$ and $L_{\tau}$ in the space and imaginary-time directions, respectively, and treat them as independent parameters.

We assume the interactions $J^{s}_{j}$ and $J^{t}_{j}$ to be independent random variables, drawn from the power-law distribution
\begin{equation}
    P(J)=\frac{1}{\Delta}J^{(-1+\frac{1}{\Delta})}~.
    \label{eq_powerlaw_disorder}
\end{equation}
with $0<J\le 1$. The parameter $\Delta$ can take values between 0 and $\infty$ and serves as a measure of the disorder strength. The clean limit (uniform $J$) is recovered when $\Delta=0$ and the distribution becomes arbitrarily broad (on a logarithmic scale) in the limit $\Delta \to \infty$.

The universal features of the phase diagram and the phase transitions are expected to be independent of the details of the disorder distributions. To test this, we have also performed some calculations with a binary disorder distribution as used in Ref.\ \cite{hrahsheh_vojta_prl_12}. We will return to this question in Sec.\ \ref{sec-5}.

%%%%%%%%%%%%%%%%%%%%%%%%%%%%%%%%%%%%%%%%%%%%%%%%%%%%%%
\section{Methods}
\label{sec-3}
%%%%%%%%%%%%%%%%%%%%%%%%%%%%%%%%%%%%%%%%%%%%%%%%%%%%%%

%%%%%%%%%%%%%%%%%%%%%%%%%%%%%%%%%%%%%%%%%%%%%%%%%%%%%%%%%%%%%%%%%%%%%%%%%%%%%%%%%%%
\subsection{Simulation Details}
\label{subsec:MC}
%%%%%%%%%%%%%%%%%%%%%%%%%%%%%%%%%%%%%%%%%%%%%%%%%%%%%%%%%%%%%%%%%%%%%%%%%%%%%%%%%%%

We use large-scale Monte Carlo simulations to study the classical Hamiltonian (\ref{eq_Hamiltonian_cl}) for  $q=6$, using interactions drawn from the distribution (\ref{eq_powerlaw_disorder}).  The simulations combine the single-spin-flip Metropolis algorithm \cite{metropolis1949monte} and the Wolff cluster algorithm \cite{wolff_prl_89}. The Wolff cluster algorithm proves beneficial near phase transitions where a single-spin-flip algorithm suffers critical slowing down, and the Metropolis algorithm
helps equilibrate small clusters of sites that are isolated from the rest of the chain by weak links. A full Monte Carlo sweep consists of one Metropolis sweep over the lattice followed by one Wolff cluster sweep (a number of cluster flips such that the total number of flipped spins equals the number of lattice sites).

The quality of the Monte Carlo equilibration is benchmarked, as usual, by comparing simulation runs with hot starts (the clock variables take random values initially) and cold starts (all clock variables take the same value initially). These tests do not show a significant dependence of the required equilibration times on the disorder strength. However, we moderately increase the number of Monte Carlo sweeps with increasing disorder strength to help reduce the additional statistical error stemming from the disorder fluctuations. Specifically, we employ  2000 equilibration sweeps and 5000 measurement sweeps for low disorder strengths, whereas we utilize 4000 equilibration sweeps and 10000 measurement sweeps for higher disorder strengths.

We simulate systems of linear sizes up to $L=320$ in the space direction and up to $L_{\tau}=15000$ in the imaginary-time direction, exploring disorder strength from $\Delta=0$ to $\Delta=2$. All results are averaged over a large number of disorder realizations,
ranging from 2000 samples for lower disorder strengths to 10000 samples for higher disorder strengths.

%%%%%%%%%%%%%%%%%%%%%%%%%%%%%%%%%%%%%%%%%%%%%%%%%%%%%%%%%%%%%%%%%%
\subsection{Observables}
\label{subsec:ob}
%%%%%%%%%%%%%%%%%%%%%%%%%%%%%%%%%%%%%%%%%%%%%%%%%%%%%%%%%%%%%%%%%%

We now turn to the observables measured in the simulations. For definiteness, we are using the language of magnetic systems.
First, we define the order parameter, the complex magnetization \cite{baek_minnhagen_pre_09}
\begin{equation}
\mathbf{M} = \frac{1}{N}\sum_{j,\tau}e^{i\theta_{j,\tau}} = |\mathbf{M}|e^{i \phi}~,
\label{eqn:comp_mag}
\end{equation}
where $\theta_{j,\tau}=2\pi p_{j,\tau} / q$ is the phase of the clock variable at the site $(j,\tau)$ and $N=L \, L_\tau$ denotes the total number of sites.
This complex magnetization is the building block of several of our primary observables. The average magnetization $m$ is given by
\begin{equation}
\label{eqn:dis_mag1}
     m= [\langle|\textbf{M}|\rangle]_{\rm dis}~.
\end{equation}
Here, the angular brackets $\langle \cdots \rangle$ indicate the thermodynamic (Monte Carlo) average,
and $[\cdots]_{\rm dis}$ denotes the average over the quenched disorder realizations.
The corresponding, disorder averaged susceptibility is
\begin{equation}
\chi = (N/T) \left[ \langle |\textbf{M}|^2 \rangle -\langle |\textbf{M}| \rangle^2 \right ]_\textrm{dis}~,
\label{eq:chi}
\end{equation}
and the Binder cumulant \cite{binder1981finite} is given by
\begin{equation}
    U_{m}=\left[1-\frac{\langle |\textbf{M}|^{4}\rangle }{2\langle |\textbf{M}|^{2}\rangle^{2}}\right]_{\rm dis}~.
    \label{eqn:binder}
\end{equation}
To identify the transition between the QLRO and TLRO phases, we also define a clock order-parameter \cite{baek_minnhagen_pre_09},
\begin{equation}
   \label{eqn:clock_op}
   m_{\phi} = \left[\langle\cos(q\phi)\rangle\right]_{\rm dis}.
\end{equation}
It vanishes in the paramagnetic phase as well as the QLRO phases, because the distribution of the global order parameter phase $\phi$ on the interval $[0,2\pi)$ is uniform. In the TLRO phase, in contrast, the system spontaneously freezes in one of the $q$ clock states which implies that the $\phi$ distribution develops maxima at multiples of $2\pi/q$, rendering $m_{\phi}$ nonzero.

In addition to these magnetization-based observables, we also measure the disorder-averaged specific heat,
\begin{equation}
\label{eq:C}
C= (N/T^{2}) \left[(\langle E^{2}\rangle-\langle E\rangle^{2})\right]_{\rm dis}~,
\end{equation}
and the correlation lengths in the spatial and imaginary-time directions. They are obtained from the second moment of the spin-spin correlation function and can be expressed as follows \cite{vojta2016quantum,COOPER1982210,kim1993application,caracciolo2001finite},
\begin{equation}
\begin{aligned}
\xi_s & =\left[\left(\frac{\tilde{G}(0,0)-\tilde{G}\left(k_{s 0}, 0\right)}{k_{s 0}^2 \tilde{G}\left(k_{s 0}, 0\right)}\right)^{1 / 2}\right]_{\mathrm{dis}} ~, \\
\xi_\tau & =\left[\left(\frac{\tilde{G}(0,0)-\tilde{G}\left(0, k_{\tau 0}\right)}{k_{\tau 0}^2 \tilde{G}\left(0, k_{\tau 0}\right)}\right)^{1 / 2}\right]_\mathrm{dis}~.
\end{aligned}
\end{equation}
Here, $\tilde{G}\left(k_{s 0}, k_{\tau 0}\right)$ is the Fourier transform of the correlation function, while $k_{s 0} =2\pi/L$ and $k_{\tau 0} = 2\pi/L_\tau$ are the minimum values of the wave numbers in space and imaginary-time directions, respectively.

%%%%%%%%%%%%%%%%%%%%%%%%%%%%%%%%%%%%%%%

Finally, we also study the helicity modulus or spin stiffness, which measures the free energy cost of twisted boundary conditions, which can be implemented by fixing the spins at two opposite boundaries at specific phases, with a fixed angle $\Theta$ between them.  For systems with a continuous $U(1)$ or $O(2)$ symmetry, the helicity modulus is usually defined as the response to an infinitesimal twist via
\begin{equation}
    \label{eq_infsm_stiffness}
    \rho_s = \left( \frac{\partial^2F}{\partial\Theta^2} \right)_{\Theta=0} L^{2-D}~,
\end{equation}
where $F$ is the total free energy.

For the $q$-state clock model, any twists of the boundary conditions are necessarily non-infinitesimal. We, therefore define the helicity modulus as the response to a finite twist $\Theta$,
\begin{equation}
\label{eq_stiffness}
    \rho_s(\Theta) = \frac{2\Delta F}{\Theta^2} L^{2-D}~,
\end{equation}
where $\Delta F = F_\Theta -F_0$ is the difference between free energies of the twisted and untwisted systems \cite{kumano_hukushima_prb_13}.
In the Monte Carlo simulations, we do not implement fixed boundary conditions but rather compare the free energies of periodic and twisted-periodic boundary conditions. For the latter, we modify the interaction across one of the boundaries to $-J^s_j \cos \left[{2\pi (p_{j,\tau}-p_{j+1,\tau})}/{q} - \Theta \right]$.

The free energy cannot be measured directly in a Monte Carlo simulation. We evaluate the free energies explicitly by integrating the (disorder averaged) internal energy $U=[\langle H_{\mathrm{cl}}\rangle]_{\mathrm{dis}}$,
\begin{equation}
    F(T)= F(T_0)+ T \int_{\beta_0}^\beta d\beta ' U(\beta ')~,
\label{eq:F_int}
\end{equation}
where $\beta=1/T$.
The starting point  $T_0$ of the integration is chosen sufficiently high, i.e., deep in a paramagnetic phase where the twist in the boundary conditions does not matter, and  $F(T_0)$ for the twisted and the untwisted systems are identical. This ensures that $F(T_0)$ drops out of the free energy difference $\Delta F$ in Eq.\ (\ref{eq_stiffness}).
For our helicity modulus calculations, we fix $T_0 =30$ (much larger than the critical temperature of the clean system) and use a temperature step $dT=0.02$ in the numerical integration (\ref{eq:F_int}). We have tested $T_0$ values as high as 50 and temperature steps as small as 0.01, with unchanged results.

To estimate the statistical error of $F$, we follow an ensemble method (see, e.g., Ref.~\cite{Khairnar2024}): After each Monte Carlo run, we generate 100 artificial internal energy data sets $U_a(T) = U(T)+ r \delta U(T)$, where $r$ is a random number taken from a normal distribution of unit variance and $\delta U(T)$ is the Monte Carlo error estimate of $U(T)$. We evaluate $F_a(T)$ for each $U_a(T)$ using the integration (\ref{eq:F_int}). The mean and standard deviation of the set of $F_a(T)$ give $\langle F(T) \rangle$ and its error $\delta F(T)$.

The helicity modulus can be used to identify the transition between the paramagnetic to QLRO phases as well as the transition between the QLRO and TLRO phases. In the paramagnetic phase, the free energy difference $\Delta F$ vanishes as $\Delta F \sim e^{-L/\xi}$, where $\xi$ is correlation length. Consequently, $\rho_s = 0$ in the thermodynamic limit. In the ordered phase, due to the presence of domain walls, $\rho_s$ diverges in the thermodynamic limit. In the intermediate QLRO phase, $\rho_s$ is expected to be nonzero and finite.

As the classical Hamiltonian (\ref{eq_Hamiltonian_cl}) is anisotropic in the presence of quenched disorder, we need to distinguish the helicity modulus $\rho_s$ for a twist in the space direction from the helicity modulus $\rho_\tau$ for a twist in the imaginary time direction. In the quantum-to-classical mapping, the latter is related to the compressibility $\kappa$ of the original quantum Hamiltonian (\ref{eqn:quantum-hamiltonian}). Correspondingly, the Luttinger parameter $\tilde{g}=\pi\sqrt{\rho_s \kappa}$ of the quantum Hamiltonian is given by $\tilde{g}= (\pi/T)\sqrt{\rho_s \rho_\tau}$ in our simulations. In the clean limit, the classical Hamiltonian is isotropic, implying $\rho_s = \rho_\tau$.

At a BKT transition, the value of the Luttinger parameter is expected to be universal. According Kosterlitz and Nelson \cite{nelson1977universal}, the Luttinger parameter at the paramagnetic-to-QLRO transition at $T_{c2}$ is given by $\tilde{g}=2$, or equivalently by $\sqrt{\rho_s\rho_\tau} = 2T_{c2}/\pi$.  Analogously, the helicity moduli at the QLRO-TLRO transition at $T_{c1}$ are expected to fulfill the relation $\sqrt{\rho_s\rho_\tau}=q^2T_{c1}/8\pi$ \cite{jose_kadanoff_prb_78}.

In our Monte Carlo calculations of the helicity modulus, we choose a twist of $\Theta=\pi$ which corresponds to anti-periodic boundary conditions. This has the advantage that the efficient Wolff cluster algorithm can be employed, whereas this would not be possible for twisted bonds  $-J^s_j \cos \left[{2\pi (p_{j,\tau}-p_{j+1,\tau})}/{q} - \Theta \right]$ with $0< \Theta < \pi$.
As pointed out in Ref.\ \cite{Khairnar2024}, at $\Theta=\pi$, the thermodynamic ensemble consists of an equal mixture of states with opposite chiralities. This leads to an additional $\ln(2)$ entropy contribution in the QLRO and TLRO phases and a corresponding correction to $F$. Including this correction in the definition (\ref{eq_stiffness}), shows that the helicity moduli at $\Theta=\pi$ are reduced by
\begin{equation}
\Delta \rho_{s,\tau} = -\frac{2\ln 2}{\pi^2} T
\label{eq:delta_rho}
\end{equation}
compared to their values for small twist angles.
This can be accounted for either by correcting the helicity modulus values arising from the simulations or
by appropriately modifying the Jose-Kadanoff and Kosterlitz-Nelson relations.

%%%%%%%%%%%%%%%%%%%%%%%%%%%%%%%%%%%%%%%%%%%%%%%%%%%%%%%%%%%%%%%%%%%%%%%%
\subsection{Finite-size scaling}
\label{subsec:FSS}
%%%%%%%%%%%%%%%%%%%%%%%%%%%%%%%%%%%%%%%%%%%%%%%%%%%%%%%%%%%%%%%%%%%%%%

In this section, we concentrate on the data analysis for the various observables introduced in Sec.~\ref{subsec:ob}
by means of finite-size scaling (FSS). As the quenched randomness breaks the symmetry between the space and imaginary time directions, the behavior close to a phase transition is governed by two independent scaling variables (dimensionless ratios between system size and correlation length), viz., $L/\xi_s$ and $L_\tau/\xi_\tau$. At a conventional critical point, the correlation lengths in both space and imaginary time are expected to diverge as powers of the distance $r= (T-T_c)/T_c$ from criticality, $\xi_s \sim |t|^{-\nu}$ and $\xi_\tau \sim \xi^z \sim |t|^{-\nu z}$, where $\nu$ and $z$ are the correlation length and dynamical critical exponents, respectively. The scaling form of dimensionless observables such as the Binder cumulant therefore reads
\begin{equation}
    \label{eqn:Binder_scal}
    U_m(r,L,L_\tau)= \tilde{U}_{m}(rL^{1/\nu},L_\tau/L^z)~.
\end{equation}
Here, $\tilde{U}_{m}$ is the scaling function. The clock order parameter $m_\phi$ at a conventional transition fulfills the same scaling form \cite{stiansen_sperstad_prb_11},
 \begin{equation}
    \label{eqn:clock_op_scal}
    m_\phi(r,L,L_\tau)= \tilde{m}_{\phi}(rL^{1/\nu},L_\tau/L^z)~.
\end{equation}

However, these scaling forms are not expected to hold at a BKT transition at which the correlation length depends exponentially on the distance from criticality, $\ln (\xi_s) \sim r^{-1/2}$ \cite{kosterlitz_thouless_jpcssp_73}. The exponential dependence is reflected in the scaling forms
\begin{eqnarray}
    \label{eqn:Binder_scal_BKT}
    U_m(r,L,L_\tau) &=& \tilde{U}_{m}(r\log(L/L_0)^2,L_\tau/L^z) ~,\\
    m_\phi(r,L,L_\tau) &=& \tilde{m}_{\phi}(r\log(L/L_0)^2,L_\tau/L^z) ~,
\end{eqnarray}
where $L_0$ is a microscopic length scale.

Certain quantum phase transitions and nonequilibrium phase transitions in disordered systems feature exotic infinite-randomness critical points. Examples include the random transverse-field Ising ferromagnet \cite{fisher_prl_92,fisher_prl_95}, the superconductor-metal quantum phase transition observed in thin nanowires \cite{hoyos_kotabage_prl_07,vojta_kotabage_prb_09} and the disordered contact process \cite{HooyberghsIgloiVanderzande03,*VojtaDickison05,*VojtaFarquharMast09}. In the infinite-randomness scenario, the power-law dependence of $\xi_\tau$ on $\xi_s$ is replaced by an exponential (activated) scaling relation $\log(\xi_{\tau})\sim \xi_{s}^\psi$, where $\psi$ is the tunneling exponent. In such a scenario the scaling forms of $U_m$ and $m_\phi$ are modified to read
\begin{eqnarray}
    U_m(r,L,L_{\tau}) &= &\tilde{U}_m(rL^\frac{1}{\nu},\log(L_\tau/L_{\tau 0})/L^{\psi})~, \label{eqn:Binder_scal_act} \\
    m_\phi(r,L,L_{\tau}) &= &\tilde{m}_\phi(rL^\frac{1}{\nu},\log(L_\tau/L_{\tau 0})/L^{\psi})~. \label{eqn:clock_op_scal_act}
\end{eqnarray}

The FSS of the clock order parameter and the Binder cumulant can be used to find the transition temperatures $T_{c1}$ and $T_{c2}$, respectively. Focusing first on the Binder cumulant, we follow the treatment elucidated in Ref.\ \cite{sknepnek_vojta_prl_04}. It is based on the fact that $U_m$ develops a maximum as a function of $L_\tau$ for fixed $L$ and $T$. The position $L_\tau^{\rm max}$ of this maximum characterizes the ``optimal'' sample shape at which the correlations extend equally in both directions. Setting $L_\tau =L_\tau^{\rm max}$ fixes the second argument of the scaling function $\tilde U_m$. Consequently, the peak value $U_{m}^{\rm max}$ is independent of the system size $L$ at criticality, $r=0$ ($T=T_{c2}$). Moreover, for the case of power-law dynamical scaling, one gets a scaling collapse of the Binder cumulant at $r=0$ as a function of $L_\tau/L_\tau^{\rm max}$. If the dynamical scaling is instead of the activated type, a data collapse obtains if $U_m$ is plotted as a function of the variable $\log{(L_\tau)}/\log{(L_\tau^{\rm max})}$. The clock order parameter can be analyzed in the same way to find the transition between the TLRO and QLRO phases at $T_{c1}$.

Further information about the phase transitions can be gained from the finite-size dependence of the disorder-averaged magnetization right at the transition point.

At a conventional phase transition, it is expected to follow a power-law form and often characterized by means of the critical exponent combination $\beta/\nu$,
\begin{equation}
     m(T_{c},L) \sim L^{-\beta/\nu}
     \label{eqn:simple_scaling_m_betanu}
\end{equation}
where $\beta$ is the order parameter exponent. $\beta/\nu$ can be connected to the anomalous (correlation function) exponent $\eta$ via a scaling relation which reads
$\eta = 2- d -z + 2\beta/\nu$ in a quantum system with $d$ space dimensions and dynamical exponent $z$. Thus, the finite-size dependence of the magnetization
can also be expressed as
\begin{equation}
     m(T_{c},L) \sim L^{-(d+z-2+\eta)/2}~.
     \label{eqn:simple_scaling_m}
\end{equation}
This form is more convenient for BKT transitions for which $\nu$ is formally infinite, and $\beta$ is not well defined as the order parameter vanishes on both sides of the transition.

For $d=1$ and $z=1$, eq.\ (\ref{eqn:simple_scaling_m}) reduces  to the familiar relation
$m \sim L^{-\frac{\eta}{2}}$ which is expected to hold in the entire QLRO phase and at the two BKT transitions at $T_{c1}$ and $T_{c2}$ \cite{baek_minnhagen_pre_09,chatelain_jsmte_14}. This allows us to extract the anomalous exponent $\eta$ at both $T_{c1}$ and $T_{c2}$ as discussed in the Ref.\ \cite{challa_landau_prb_86}. In the absence of disorder, $\eta$ is known to take the value $1/4$ at the transition between the QLRO and paramagnetic phases at $T_{c2}$. On the other hand, the theoretical estimate of the exponent at the transition between the TLRO and the QLRO transition is $\eta=4/q^{2}$.

%%%%%%%%%%%%%%%%%%%%%%%%%%%%%%%%%%%%%%%%%%%%%%%%%%%%%%%%%%%%%%%
\section{Results and Discussions}
\label{sec-4}
%%%%%%%%%%%%%%%%%%%%%%%%%%%%%%%%%%%%%%%%%%%%%%%%%%%%%%%%%%%%%%%%%%%%

In this section, we apply our simulation and data analysis techniques to the classical Hamiltonian (\ref{eq_Hamiltonian_cl}). We first study the clean, translationally invariant case to test our methods. In Sec.~\ref{subsec:4-b}, we establish the phase diagram of the disordered model. Finally,  Sec.~\ref{subsec:4-c} and Sec.~\ref{subsec:4-d} are devoted to the study of the critical behavior of the model in the weak and strong disorder regimes, respectively.

%%%%%%%%%%%%%%%%%%%%%%%%%%%%%%%%%%%%%%%%%%%%%%%%%%%%%%%%%%%%%%%%
\subsection{Clean Case}
\label{subsec-4a}
%%%%%%%%%%%%%%%%%%%%%%%%%%%%%%%%%%%%%%%%%%%%%%%%%%%%%%%%%%%%%%%%%

\begin{figure}
  \includegraphics[width=\columnwidth]{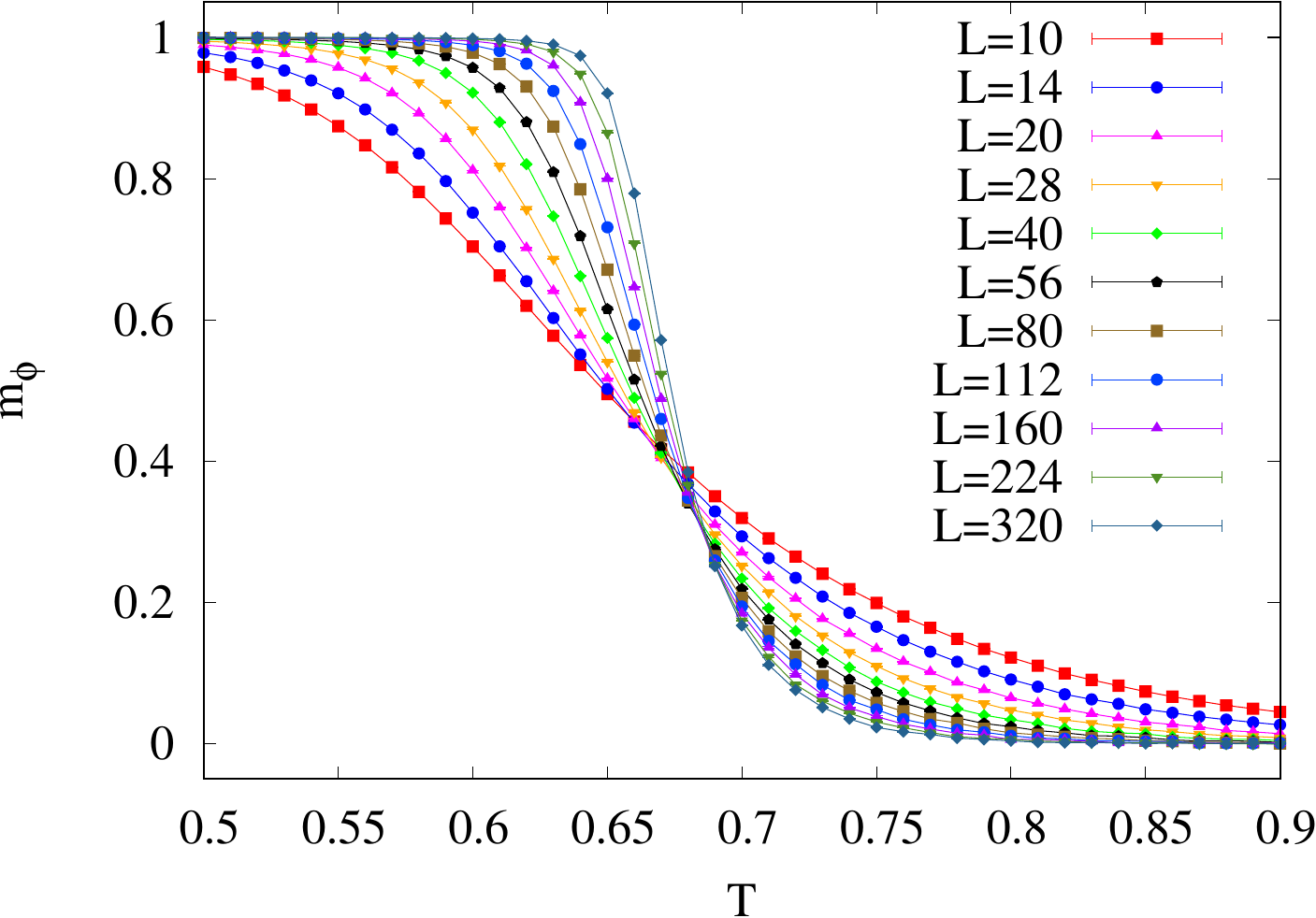}
  \includegraphics[width=\columnwidth]{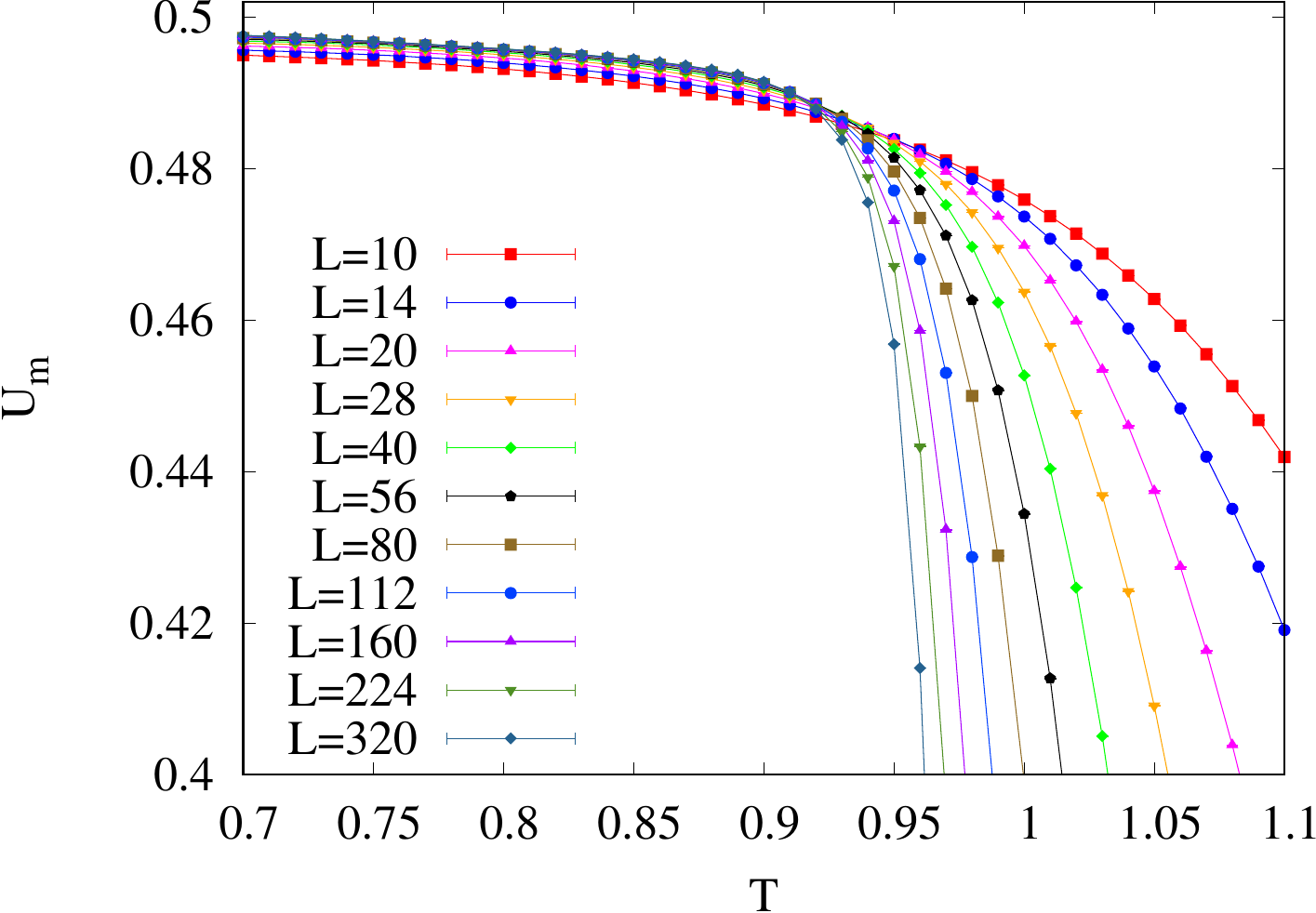}
    \caption{Clock order parameter $m_{\phi}$ and Binder cumulant $U_m$ of the clean system ($\Delta=0$) as functions of temperature $T$ for several linear system sizes $L$. The statistical errors are smaller than the symbol size. The critical temperatures $T_{c1}, T_{c2}$ can be determined from intersections of $m_\phi$ and $U_m$ curves, respectively. Corrections to finite-size scaling systematically shift the intersection point $T^*$.}
    \label{fig:crossing_op}
\end{figure}

\begin{figure}
 \includegraphics[width=\columnwidth]{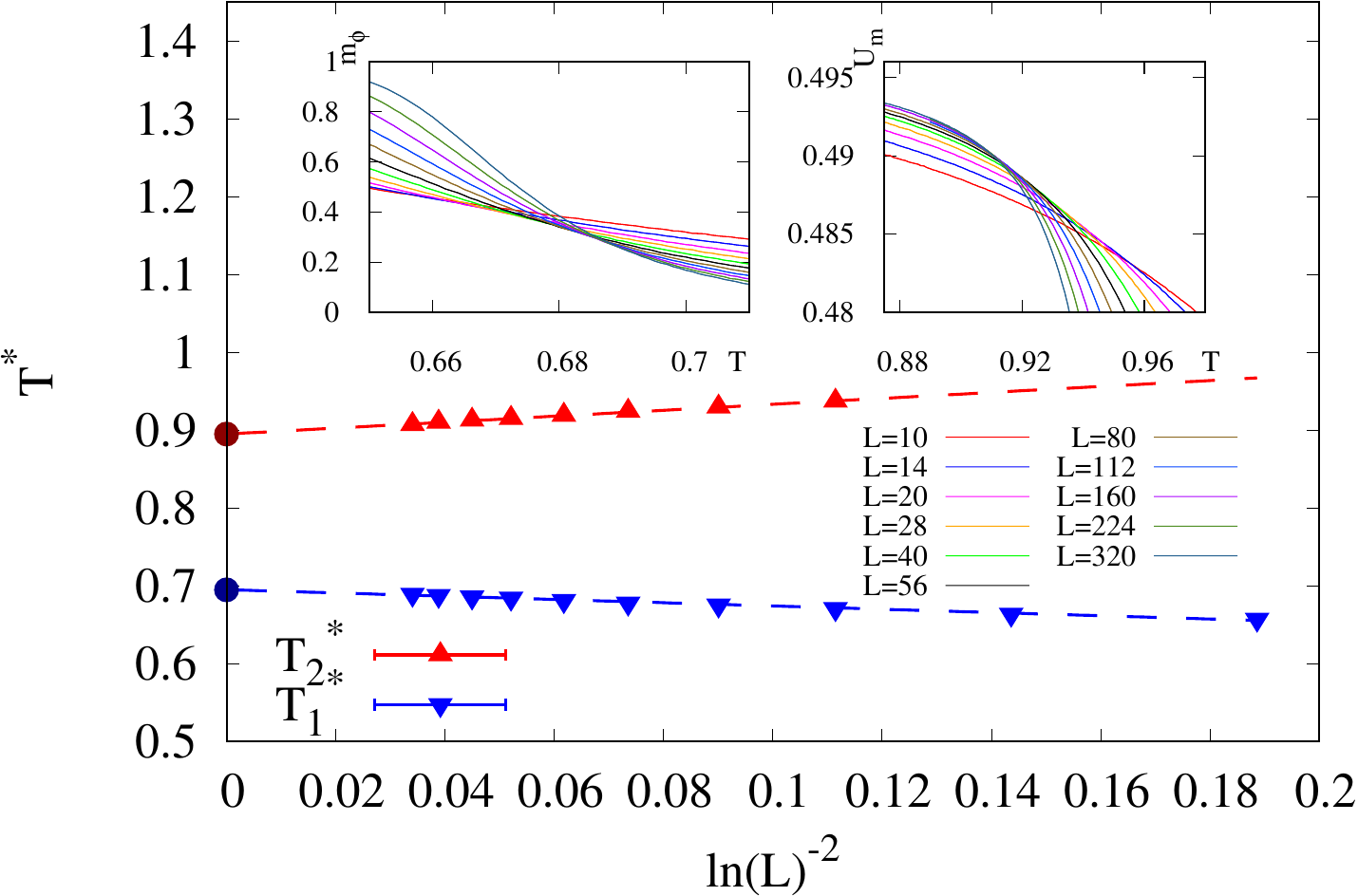}
    \caption{Extrapolation of the crossing temperatures $T^*(L)$ of $m_{\phi}$ (bottom, blue) and $U_{m}$  (top, red) in the clean case to infinite system size. The transition temperatures are obtained as the intercepts on the y-axis from fits using $T^*(L) = T_c + A (\ln L)^{-2}$, yielding $T_{c1}=0.695(8)$  and $T_{c2}=0.895(8)$.
    The insets show $m_{\phi}$ (left) and $U_{m}$ (right) close to the respective transitions.}
    \label{fig:extrapolation_tcs_clean}
\end{figure}

In order to identify the phase transitions of the clean ($\Delta=0$) 6-state clock Hamiltonian (\ref{eq_Hamiltonian_cl}), we analyze the clock order parameter $m_{\phi}$ and the Binder cumulant $U_m$. Figure \ref{fig:crossing_op} depicts $m_{\phi}$ and $U_m$ as functions of the temperature $T$ for different system sizes $L$.
As the symmetry between the space and imaginary time directions is not broken in the clean case, fixing the dynamical exponent at $z=1$, we use square, $L \times L$, samples.
To extract the transition temperatures $T_{c1}$, and $T_{c2}$, we find the crossing point $T^{*}(L)$ of the curves of the relevant observable at two different system sizes, $L$ and $aL$, where $a$ is a constant. (We employ $a = \sqrt{2}$ in the following.) The crossing points shift as a function of system size because of corrections to scaling. At a BKT transition, this shift is expected to follow the functional form \cite{baek_minnhagen_pre_09}
\begin{equation}
\label{eq_Tc_extrapolate}
    T^*(L)-T_c \sim  (\ln L)^{-2} ~.
\end{equation}
Here, $T_{c}$ is the corresponding asymptotic critical temperature ($T_{c1}$ for the crossings of $m_\phi$ and $T_{c2}$ for the crossings of $U_m$). The extrapolations of $T^*(L)$ to infinite system size are shown in Fig.~\ref{fig:extrapolation_tcs_clean}, they yield $T_{c1}=0.695(8)$  and $T_{c2}=0.895(8)$.

\begin{figure}
    \includegraphics[width=\columnwidth]{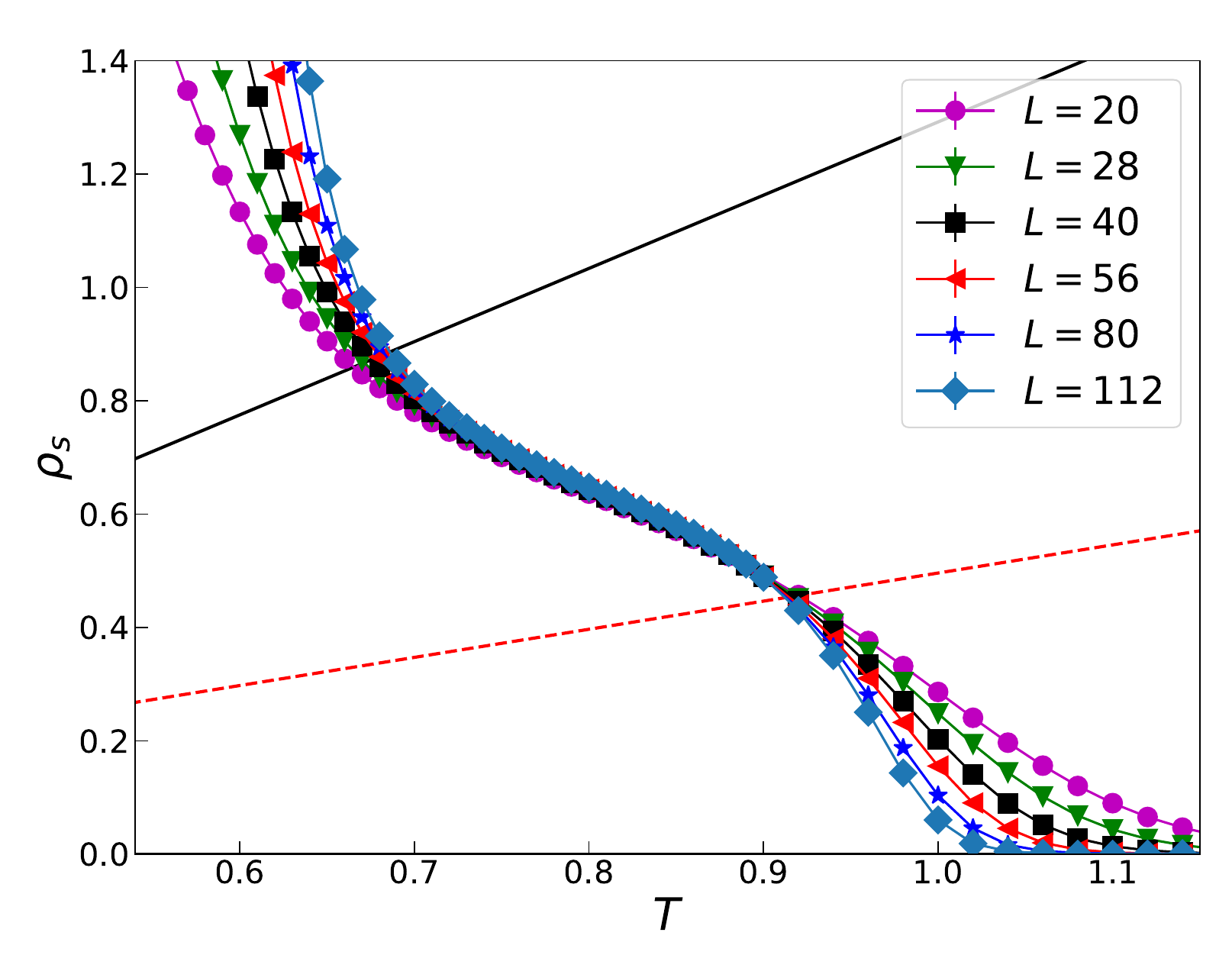}
    \caption{Helicity modulus $\rho_s$ for twist angle $\Theta=\pi$ vs.\ temperature $T$ for different linear system sizes $L$ in the clean case, $\Delta=0$. The solid (black) line corresponds to the modified Jose-Kadanoff relation $\rho_{s}=T(q^{2}/8\pi-2\ln 2/\pi^2)$, and the dashed (red) line depicts the modified Kosterlitz-Nelson relation $\rho_{s}=T(2/\pi -2\ln 2/\pi^2)$. The critical temperatures are obtained by extrapolating the crossing temperatures $T^{*}(L)$ between these lines and the data to $L\to\infty$ according to the BKT ansatz (\ref{eq_Tc_extrapolate}). We find $T_{c1}=0.707(2)$ and $T_{c2}=0.895(6)$.}
    \label{fig:stiffness_D000}
\end{figure}

Alternatively, we can lay recourse to the helicity modulus $\rho_s$ to extract the transition temperatures $T_{c1}$ and $T_{c2}$. As the clean system is isotropic, implying $\rho_\tau=\rho_s$, we only compute the spatial helicity modulus $\rho_s$.
Our simulation results for $\rho_s(\pi)$ are shown in Fig.\  \ref{fig:stiffness_D000} as functions of temperature for various system sizes.
Because we employ a twist of $\Theta=\pi$, we use the modified Jose-Kadanoff and Kosterlitz-Nelson relations, as discussed in Sec.\ \ref{subsec:ob}. Due to the presence of finite size corrections, these lines do not intersect with the helicity modulus curves at a single point. Instead, there is a systematic shift of the intersection points with $L$. The critical temperatures are obtained by extrapolating the intersection temperatures $T^{*}(L)$ to the thermodynamic limit according to Eq.\ (\ref{eq_Tc_extrapolate}). From the fits, we obtain $T_{c1}=0.707(2)$ and $T_{c2}=0.895(6)$, which agrees with the results from $m_\phi$ and $U_m$. Moreover, these transition temperatures also agree reasonably well with recent high-accuracy results for the clean 6-state clock model in the literature, see, e.g., Table I in Ref.\ \cite{LiPaiGu22} \footnote{Some of the recent studies of the clean 6-state clock model quote very tight error bars for $T_{c1}$ and $T_{c2}$. Note, however, that the error bars of different results often do not overlap, casting some doubt on the precision of at least some of the estimates.}

%%%%%%%%%%%%%%%%%%%%%%%%%%%%%%%%%%%%%%%%%%%%%%%%%%%%%%%%%%%%%
\subsection{Disordered case: Phase diagram}
\label{subsec:4-b}
%%%%%%%%%%%%%%%%%%%%%%%%%%%%%%%%%%%%%%%%%%%%%%%%%%%%%%%%%%%%%

\begin{figure}
 \includegraphics[width=\columnwidth]{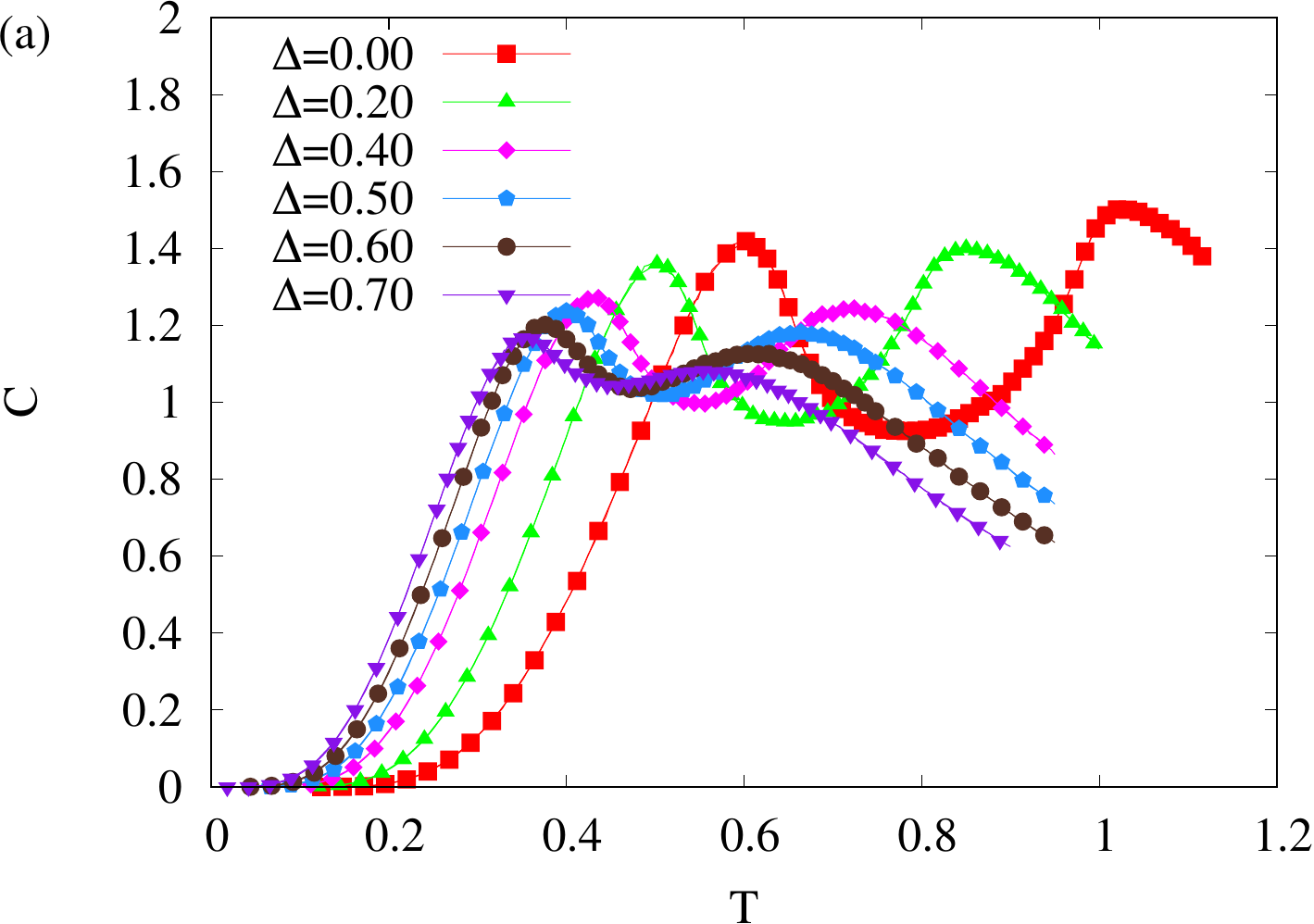}
 \includegraphics[width=\columnwidth]{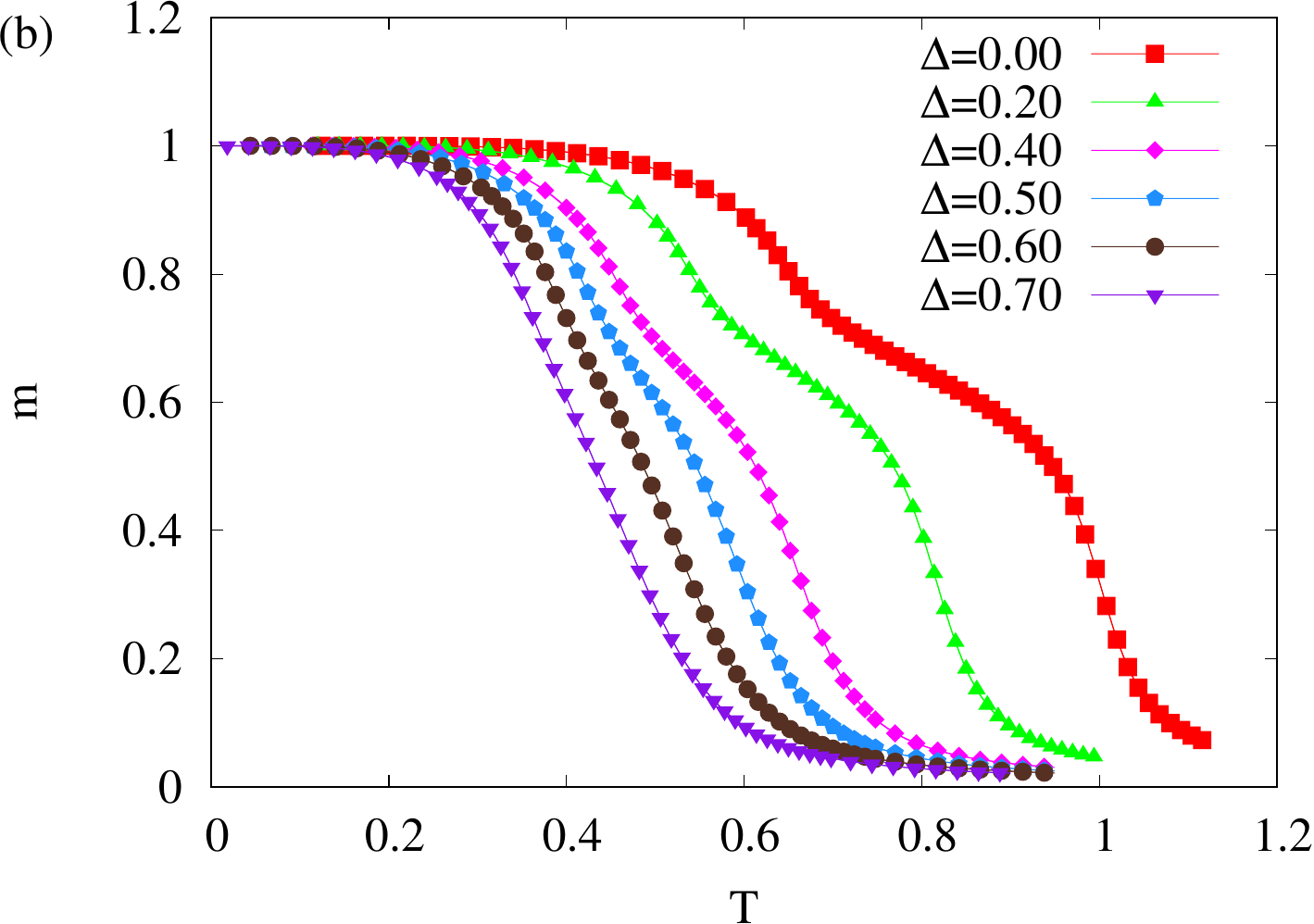}
        \caption{Specific heat $C$ (a) and magnetization $m$ (b) vs.\ temperature $T$ for different disorder strengths $\Delta$ and linear system size $L=L_\tau=128$. The temperature region between the two peaks gets narrower as the disorder increases, and the shoulder in $m(T)$ vanishes.}
    \label{fig:observables_q_6}
\end{figure}

We now turn our attention to the disordered clock model ($\Delta > 0$). We first study the specific heat as it is sensitive to all phase transitions in the system. Figure \ref{fig:observables_q_6}(a) presents the specific heat $C$ for various disorder strengths as a function of temperature for a fixed system of size $L=L_\tau=128$.
In the clean limit, $\Delta=0$, the specific heat clearly displays a double peak structure indicative of two phase transitions at $T_{c1}$ and $T_{c2}$, identified in Sec.\  \ref{subsec-4a}.

Note that the specific heat at a BKT transition features only a weak (unobservable) essential singularity right at $T_c$, but a broad peak appears slightly above $T_c$ due to the entropy release from the vortex-pair unbinding. Thus, the peak positions in Fig.\ \ref{fig:observables_q_6}(a) do not exactly coincide with the transition temperatures. The BKT character of the transition  also implies that the specific heat shows only weak system-size  effects. We have explicitly confirmed that the specific heat for  $L=L_\tau=256$ is almost identical to the result for $L=L_\tau=128$ presented in the figure. The absence of a specific heat divergence at $T_c$ is in line with the Harris criterion \cite{harris_jpcssp_74}, as will be discussed in more detail in Sec.\ \ref{subsec:4-c}.

With increasing disorder strength $\Delta$, the two peaks in the specific heat curves come closer to each other, indicating the shrinking and potential destruction of the QLRO phase.  The behavior of the specific heat is mirrored by the behavior of the magnetization $m$ in Fig.\ \ref{fig:observables_q_6}(b): For zero and weak disorder, we see a shoulder-like feature indicative of the intermediate QLRO phase \cite{lappili_pfeifer_prl_06}. This feature shrinks and disappears with increasing disorder strength, indicating the loss of the QLRO phase.

\begin{figure}
    \includegraphics[width=\columnwidth]{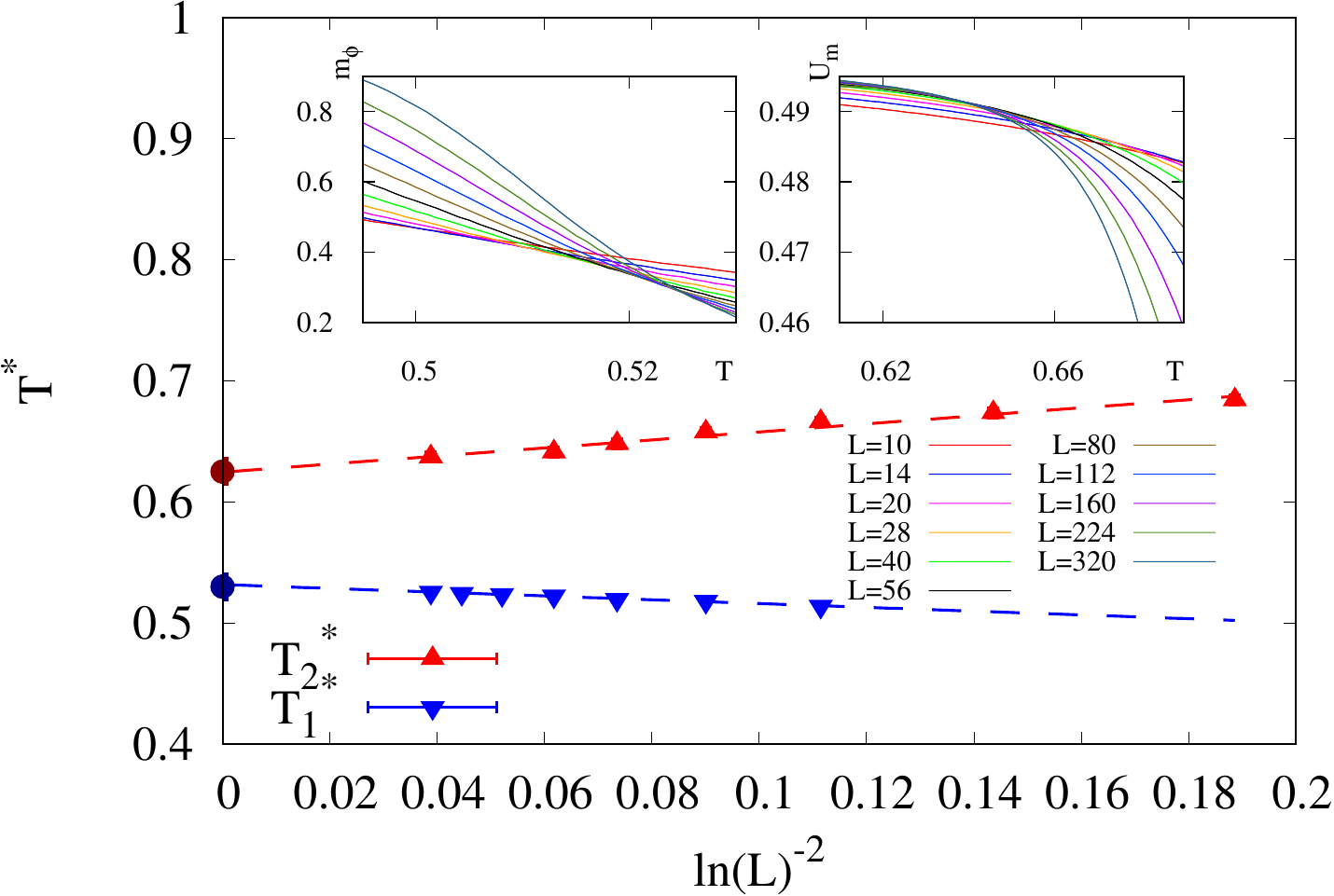}
    \caption{Extrapolation of crossing temperatures $T^*(L)$ of $m_{\phi}$ (bottom, blue) and $U_{m}$  (top, red)  for different system sizes and disorder strength $\Delta=0.3$. The transition temperatures are obtained as the $y$-intercepts from fits using $T^*(L) = T_c + A (\ln L)^{-2}$. These yield $T_{c1}=0.530(5)$  and $T_{c2}=0.625(10)$.
    The insets show $m_{\phi}$ (left) and $U_{m}$ (right) close to the respective transitions. }
    \label{fig:extrapolation_tcs}
\end{figure}

\begin{figure}
    \includegraphics[width=\columnwidth]{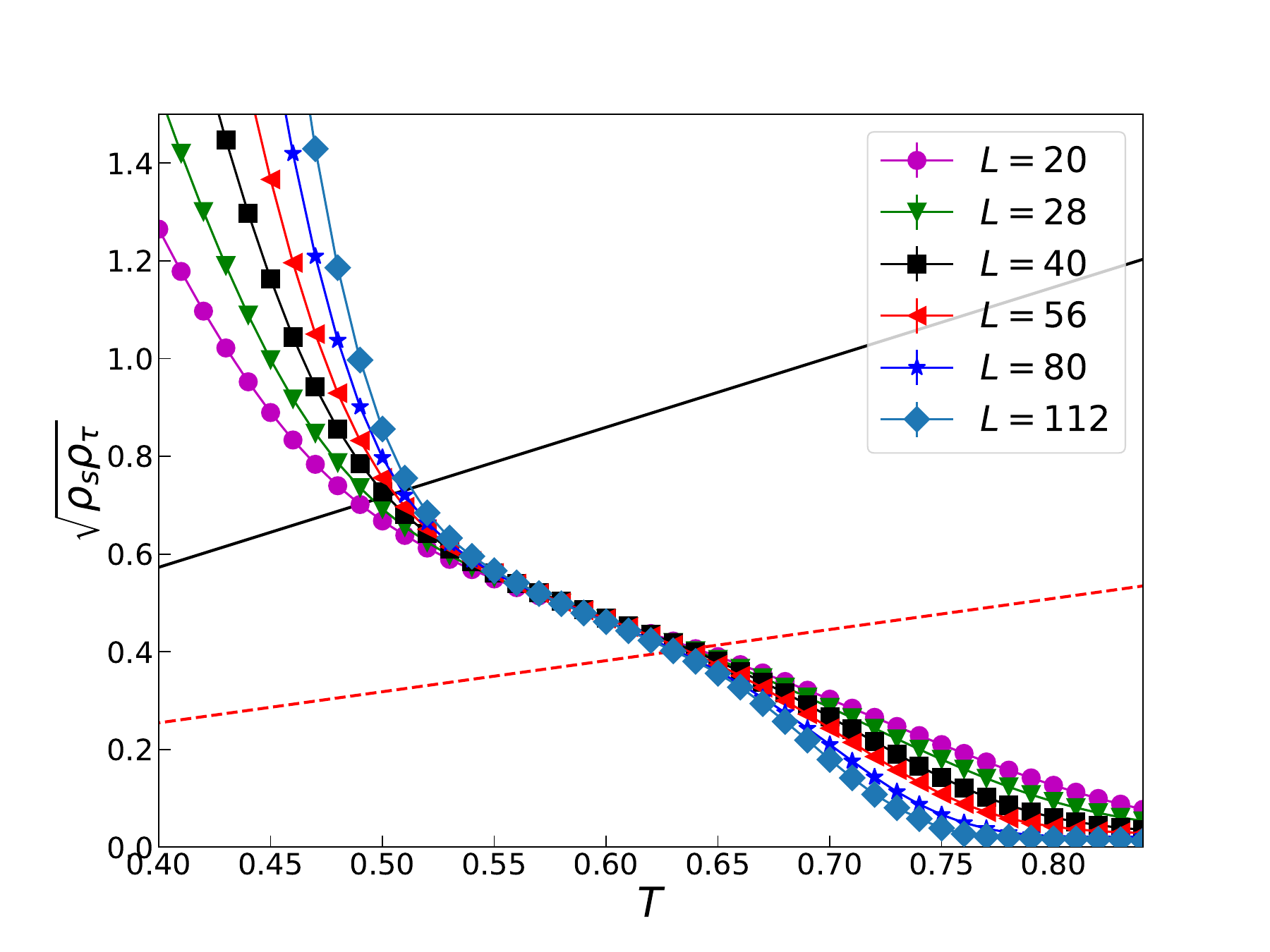}
    \caption{Helicity modulus combination $\sqrt{\rho_s\rho_\tau}$ vs. temperature $T$ for different system sizes $L$ at disorder strength $\Delta=0.3$. Both $\rho_s$ and $\rho_\tau$ have been corrected according to Eq.\ (\ref{eq:delta_rho}). The solid (black) line shows the Jose-Kadanoff relation $\sqrt{\rho_s\rho_\tau}=T q^{2}/8\pi$, and the dashed (red) line depicts the  Kosterlitz-Nelson relation $\sqrt{\rho_s\rho_\tau}= 2 T/\pi$. $T_{c1}$ and $T_{c2}$ are determined by extrapolating the intersections of these lines with the data according to Eq.\ (\ref{eq_Tc_extrapolate}), resulting in $T_{c1}=0.53(1)$ and $T_{c2}=0.63(1)$.}
    \label{fig:stiffness_D030}
\end{figure}

To construct the temperature-disorder phase diagram of the disordered clock model  (\ref{eq_Hamiltonian_cl}) quantitatively,
we lay recourse to the methods employed for the clean case in Sec.~\ref{subsec-4a}, i.e., we determine the phase transition temperatures from the behavior of $m_\phi$, $U_m$, $\rho_s$, and $\rho_\tau$. A prototypical example for the analysis of $m_\phi$ and $U_m$ in the disordered case is presented in Fig.~\ref{fig:extrapolation_tcs} for $\Delta =0.3$.
It gives the transition temperatures $T_{c1}= 0.530(5)$ and $T_{c2}=0.625(10)$.
The helicity moduli can also be used to glean $T_{c1}$, and $T_{c2}$. In the presence
of disorder, $\rho_s$ and $\rho_\tau$ are not identical anymore. As discussed in Sec.\ \ref{subsec:ob}, we therefore need to analyze the Luttinger parameter or, equivalently, the geometric mean $\sqrt{\rho_s\rho_\tau}$.
A representative example for the same disorder strength $\Delta=0.3$ is shown in Fig.~\ref{fig:stiffness_D030}.
We find $T_{c1}=0.53(1)$ and $T_{c2}=0.63(1)$ for $\Delta=0.3$ from the extrapolations using Eq.\ (\ref{eq_Tc_extrapolate}). These values agree with those obtained from the clock order parameter $m_{\phi}$ and the Binder cumulant $U_m$ within their error bars.

\begin{figure}
    \includegraphics[width=\columnwidth]{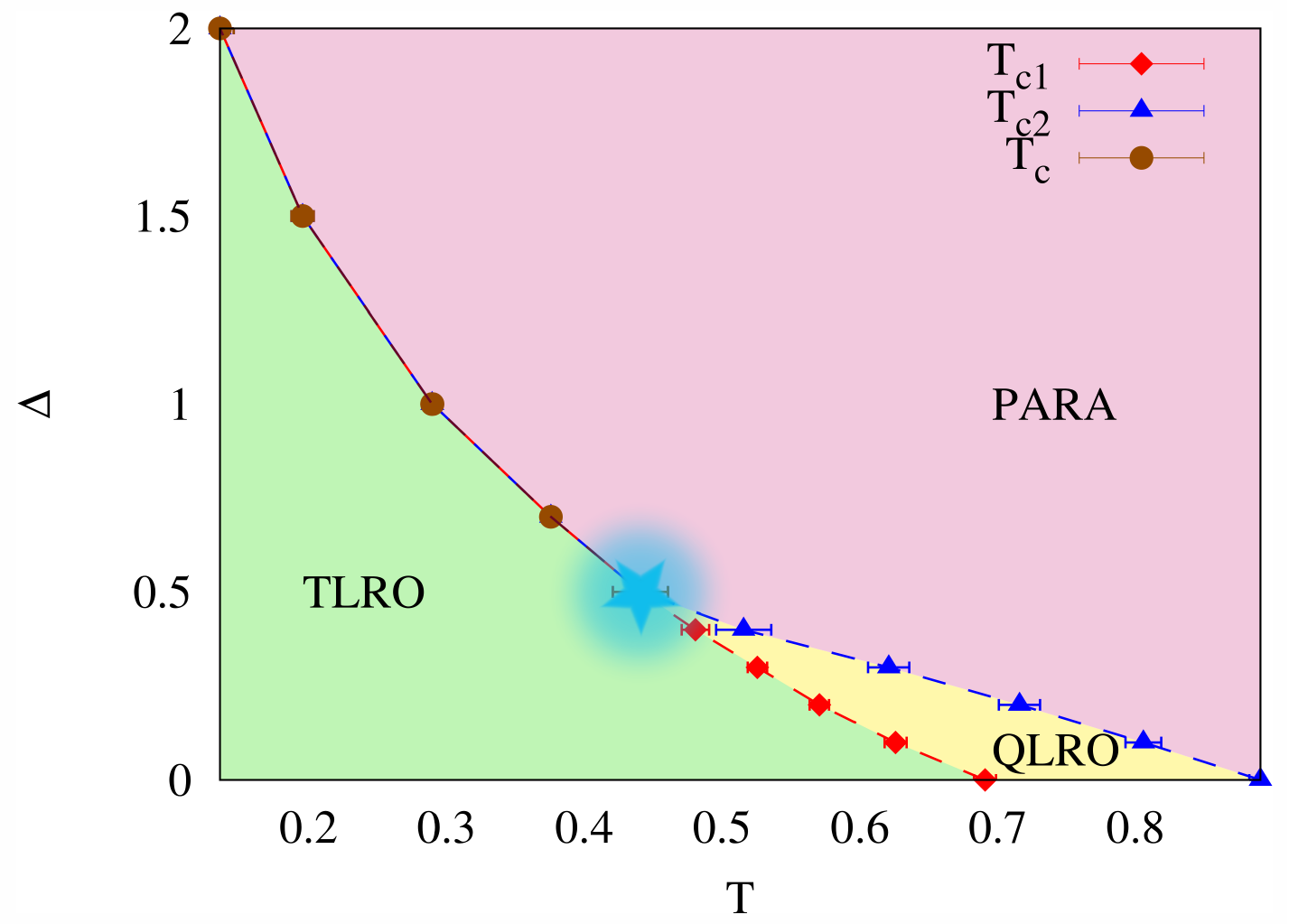}
    \caption{Phase diagram of the disordered $6$-state clock model (\ref{eq_Hamiltonian_cl}).  The transition between the TLRO and QLRO phases occurs at $T_{c1}$ (red squares), the transition between the QLRO and paramagnetic phase occurs at $T_{c2}$ (blue triangles), and the direct transition between the TLRO and paramagnetic phases is denoted by $T_{c}$ (brown circles). The multicritical point is located somewhere in the shaded region, near $\Delta=0.5$, where $T_{c1}$ and $T_{c2}$ merge. The lines are guides to the eye only.}
    \label{fig:phase-diagram}
\end{figure}

The full phase diagram is obtained by repeating these procedures for various disorder strengths. For weak disorder up to $\Delta=0.5$, we use square samples $L=L_\tau$. In the context of finite-size scaling, this assumes a dynamical exponent of $z=1$ which will be confirmed in Sec.\ \ref{subsec:4-c}. For stronger disorder, we perform the full anisotropic scaling analysis discussed in Sec.\ \ref{subsec:FSS} to determine the optimal shapes and the corresponding dynamical or tunneling exponent (see also Sec.\ \ref{subsec:4-d}). The phase diagram resulting from this analysis is shown in Fig.~\ref{fig:phase-diagram}.
In line with our expectations gleaned from the study of the specific heat $C$ and the magnetization $m$, we see that the intermediate QLRO phase shrinks with increasing disorder strength $\Delta$. This means, the phase transitions at $T_{c1}$ and $T_{c2}$ come closer to each other as the $\Delta$ increases and finally merge at a multicritical point
\footnote{Thermodynamic considerations can restrict the possible topologies of phased diagrams with multicritical points \cite{yip1991thermodynamic}. In the case of the disordered clock model, the weak, essential singularity of the free energy at the BKT transitions at $T_{c1}$ and $T_{c2}$  makes the multicritical point thermodynamically possible.}.
We estimate the multicritical point to be located at $\Delta \approx 0.5$; a more accurate determination is hampered by strong finite-size effects due to the crossovers between the different critical behaviors.
Qualitatively, the phase diagram matches with the expectation sketched in Fig.\ \ref{fig:Senthil-phasediagram}.

%%%%%%%%%%%%%%%%%%%%%%%%%%%%%%%%%%%%%%%%%%%%%%%%%%%%%%%%%%%%%%%%%%%
\subsection{Critical behavior in the weak-disorder regime}
\label{subsec:4-c}
%%%%%%%%%%%%%%%%%%%%%%%%%%%%%%%%%%%%%%%%%%%%%%%%%%%%%%%%%%%%%%%%%%%

The correlation length exponent $\nu$ of a BKT transition is formally infinite and thus fulfills the Harris criterion $d\nu > 2$ \cite{harris_jpcssp_74}.

Correspondingly, the specific heat exponent $\alpha$, obtained from the scaling relation $2-\alpha = d\nu$ is formally $-\infty$, in agreement with the missing specific heat divergence discussed after Fig.\ \ref{fig:observables_q_6}.
The Harris criterion thus implies that the critical behaviors of both transitions are expected to agree with the corresponding clean critical behaviors for sufficiently weak disorder.

However, the Harris criterion does not forbid a change of critical behavior for stronger disorder.

\begin{figure}
    \includegraphics[width=\columnwidth]{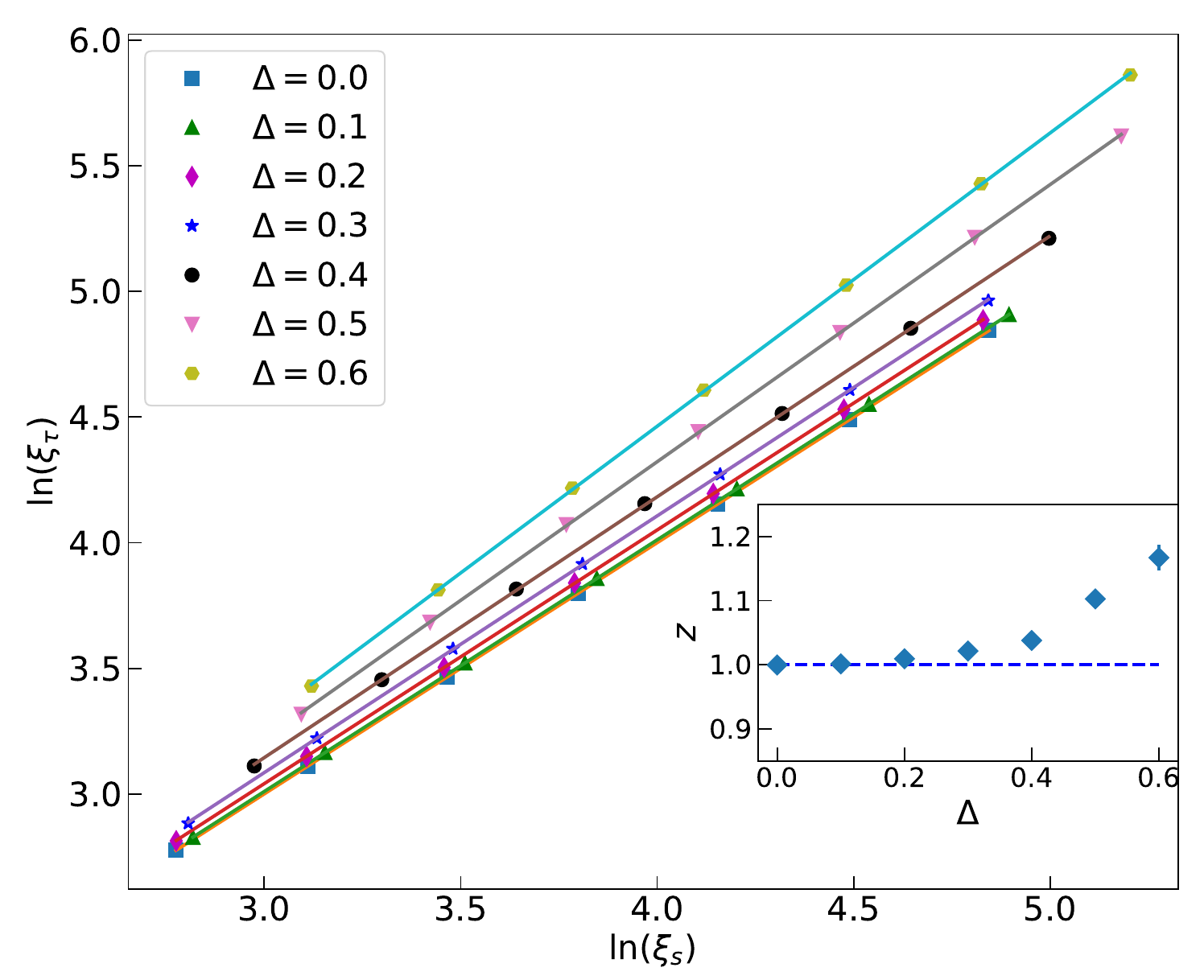}
    \caption{Correlation length $\xi_{s}$ in space direction vs.\ correlation length $\xi_{\tau}$ in imaginary-time direction near $T_{c2}$ for several disorder strengths $\Delta$.
    The solid lines are fits to $\xi_{\tau}= A \xi_s^z$. Inset: Estimate of dynamical exponent $z$ as a function of $\Delta$.
    The dashed line denotes the clean critical value $z=1$}
    \label{fig:correlation_length_vs_disorder}
\end{figure}

We start by analyzing the dynamical scaling at the transition between the QLRO and paramagnetic phases. To this end, we study the relation between the spatial correlation length $\xi_s$ and the imaginary time correlation length $\xi_\tau$.
Figure \ref{fig:correlation_length_vs_disorder} presents $\xi_{\tau}$ as a function of $\xi_s$  for various disorder strengths
at $T_{c2}$.
The data follow the power-law form $\xi_\tau \sim \xi_s^z$ expected for conventional dynamical scaling. The estimates of the dynamical exponent $z$ that results from fits to this functional form are shown in the inset of the figure. For all disorder strengths below the putative multicritical disorder $\Delta\approx 0.5$, the $z$ values remain at or very close to the clean value $z=1$. We believe that the small deviations can be attributed to finite-size effects in combination with the crossover induced by the multicritical point
\footnote{The finite size effects stem, at least partially, from the fact that $\xi_s$ and $\xi_\tau$ are finite-size correlation lengths. Extrapolating them to infinite system size should mitigate the finite-size effects.  We have attempted to estimate the bulk correlation lengths, $\xi_s^{\infty}, \xi_{\tau}^{\infty}$, following the protocol described in Ref.\ \cite{kim1994asymptotic}. Unfortunately, the accumulation of errors during this procedure demands very high accuracy input data beyond our numerical capabilities.}.
The fact that the dynamical exponent remains at $z=1$ at the transition between the QLRO and paramagnetic phases agrees with the analogous result for the disordered $(1+1)$-dimensional quantum rotor model \cite{hrahsheh_vojta_prl_12}. The fitted values of $z$ for $\Delta\gtrapprox 0.5$ need to be understood as effective exponents as the behavior for these disorder strengths is strongly affected by the multi-critical point and the crossover between the weak and strong disorder regimes.

\begin{figure}%[H]
        \centering
       \includegraphics[width=\columnwidth]{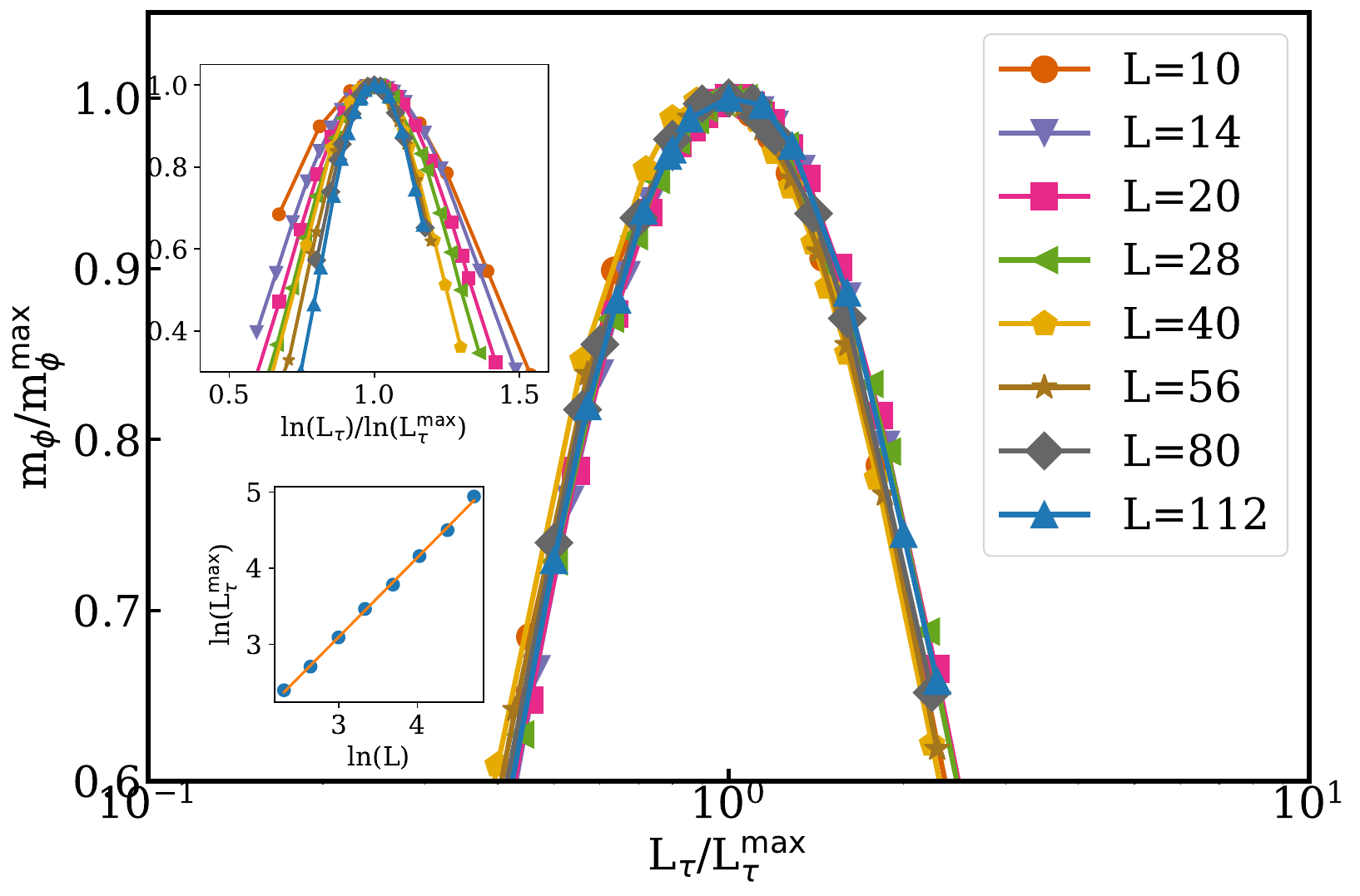}
    \caption{Anisotropic finite-size scaling of the clock order parameter $m_\phi$ near $T_{c1}\approx 0.485$ at $\Delta = 0.4$ using the power-law (conventional) scaling ansatz. Top inset: Same data analyzed using the activated scaling ansatz. Bottom inset:  $\ln(L)$ vs $\ln(L_{\tau}^{\rm max})$, the straight line represents a fit with  $L_{\tau}^{\rm max}= A L^z$. It yields $z=1.04(2)$.}
    \label{fig:scalings_delta_0.4}
    \end{figure}

To investigate the dynamical scaling at the transition between the TLRO and QLRO phases, we perform the anisotropic
finite-size scaling analysis of the clock order parameter $m_\phi$, as discussed in Sec.\ \ref{subsec:FSS}. This means we compute $m_\phi$ as a function of $L_\tau$ for fixed $L$ right at the critical temperature $T_{c1}$. A prototypical scaling plot of $m_{\phi}/m_{\phi}^{\rm max}$ vs.\ $L_{\tau}/L_{\tau}^{\rm max}$ for disorder strength $\Delta=0.4$ is presented in
Fig.~\ref{fig:scalings_delta_0.4}.
The data for other disorder strengths below the multicritical point are expected to behave analogously.
As is evident from the figure, the scaling collapse in accordance with the conventional scaling ansatz (\ref{eqn:clock_op_scal}) is very good, whereas the data do not collapse when analyzed according to the activated scaling ansatz (\ref{eqn:clock_op_scal_act}), as shown in the top inset. This demonstrates that the dynamical scaling is of a power-law type. An estimate of the the dynamical exponent $z$ can be extracted from fitting the $L_\tau^{\rm max}$ vs.\ $L$ data shown in the bottom inset in Fig.\ \ref{fig:scalings_delta_0.4} with $L_\tau^{\rm max} =a L^z$. This again yields a value very close to $z=1$. As above, we believe that the small deviation from unity is caused by finite-size effects. We thus conclude that the dynamical scaling at both transitions below the multicritical disorder strength is of conventional power law type, and the (asymptotic) dynamical exponent retains the value $z=1$, justifying the use of square samples in the rest of the simulations in the weak-disorder regime.

\begin{figure}%
        \includegraphics[width=\columnwidth]{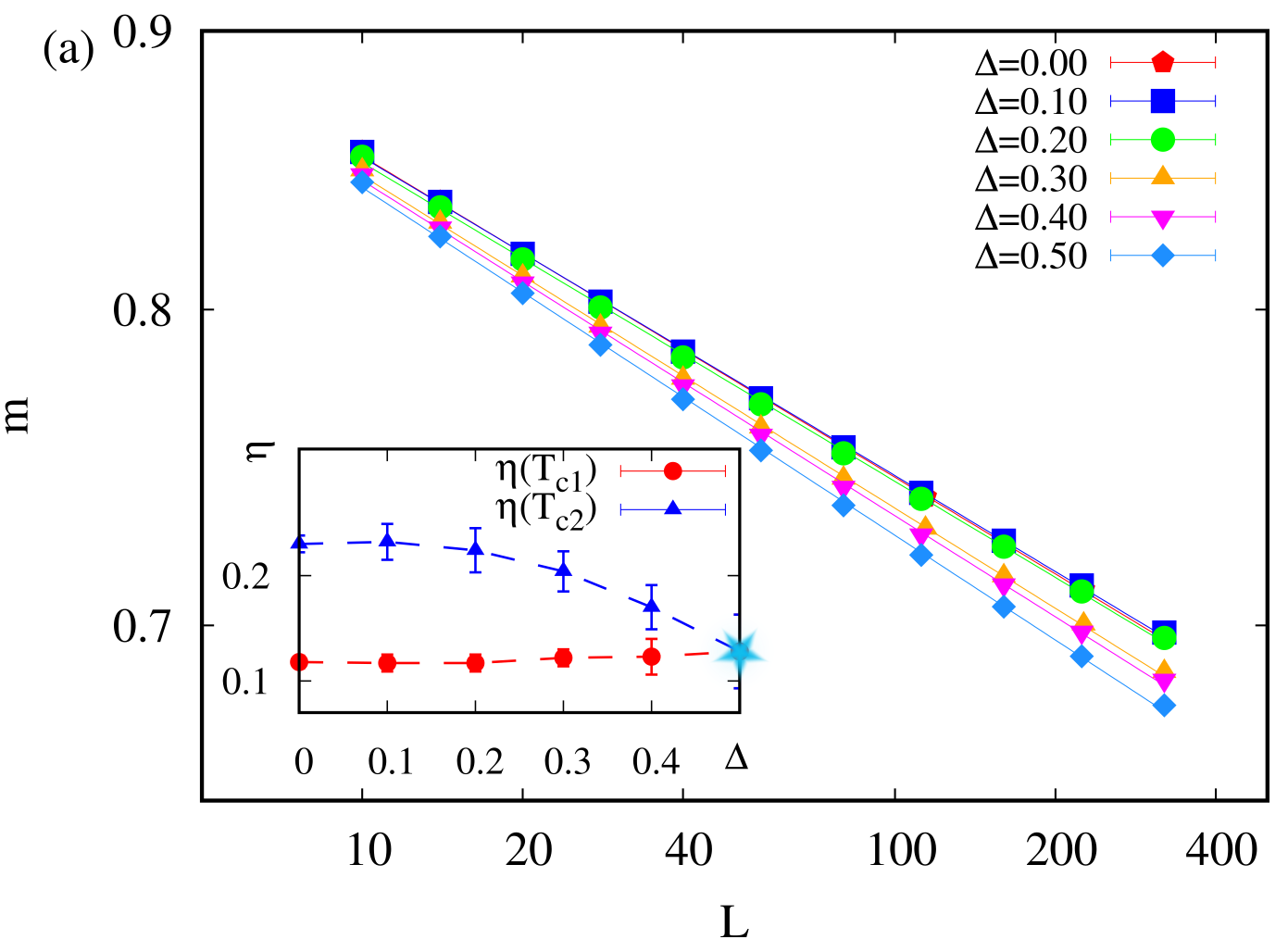}
        \includegraphics[width=\columnwidth]{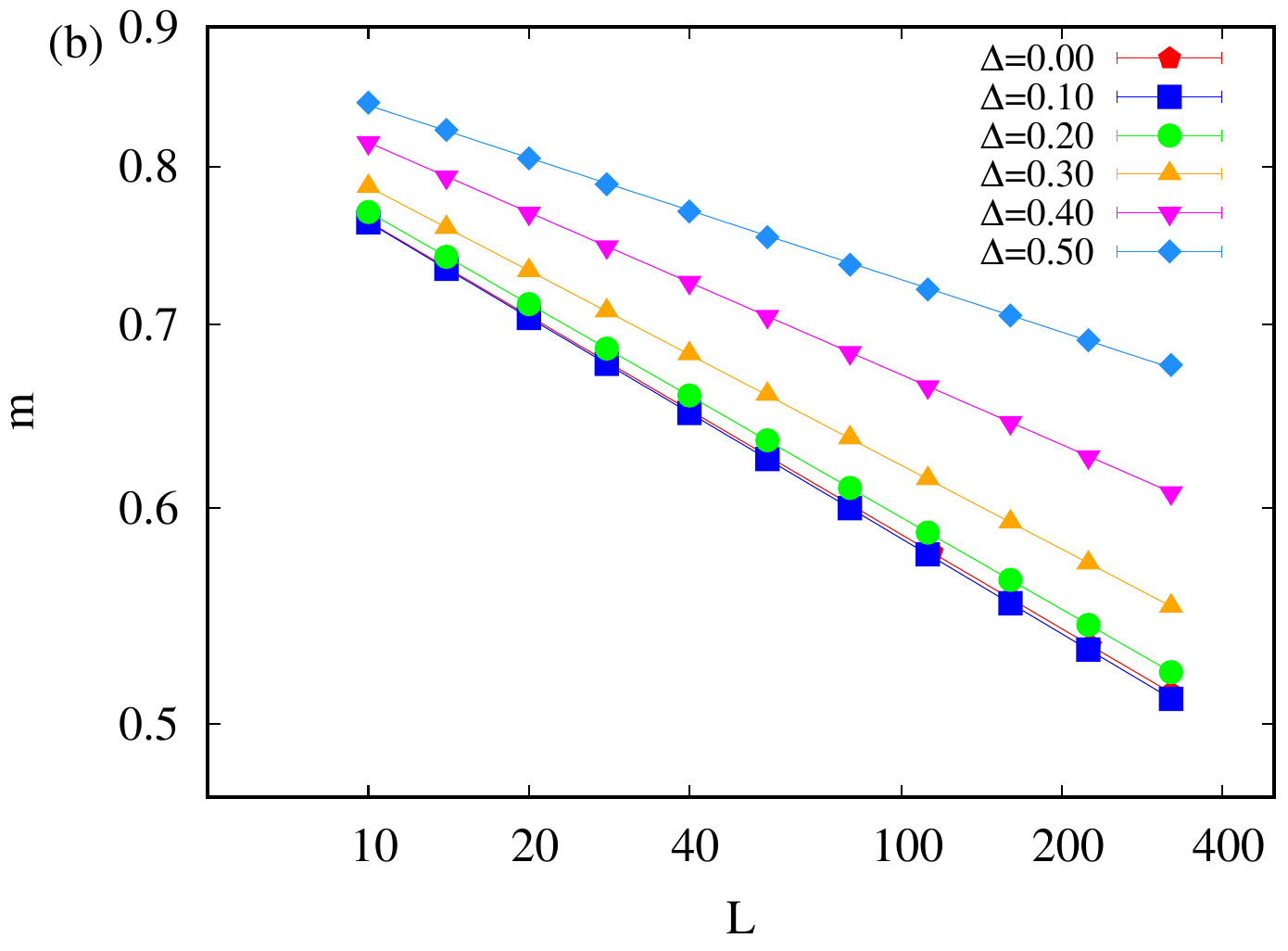}
    \caption{Magnetization $m$ vs.\ system size $L$ at (a) $T_{c1}$ and (b) $T_{c2}$.
    The solid lines are fits to the relation $m\sim L^{-\eta/2}$. Inset: Exponents $\eta(T_{c1})$ and $\eta(T_{c2})$ vs.\ disorder strength $\Delta$. The shaded region close to $\Delta=0.5$ marks the location of the multicritical point. The lines are guides to the eye only. }
    \label{fig:fit_m}
\end{figure}

We now turn to measuring the anomalous exponent $\eta$. Figures \ref{fig:fit_m}(a) and \ref{fig:fit_m}(b) show the dependence of the magnetization $m$ on the system size at $T_{c1}$ and $T_{c2}$, respectively, for a range of disorder strengths.
The data at both transitions agree very well with the expected power-law form (\ref{eqn:simple_scaling_m}). Using
$d=z=1$, Eq.\ (\ref{eqn:simple_scaling_m}) turns into the relation $m\sim L^{-\eta/2}$. The anomalous exponents $\eta$ are extracted from fits of the data with this functional form, and their values are shown in the inset of Fig.\ \ref{fig:fit_m}(a)
as functions of the disorder strength $\Delta$. The behavior of $\eta$ at the QLRO-to-paramagnetic transition is interesting.
As the disorder strength $\Delta$ is increased from zero, $\eta$ initially remains at its clean value (which is somewhat below the theoretical expectation of 1/4, likely due to finite-size effects). For $\Delta \gtrapprox 0.3$, in contrast, $\eta$ decreases significantly and displays a prominent disorder dependence. This decrease in $\eta$ is reminiscent of the behavior observed in a disordered $(1+1)$-dimensional XY model \cite{hrahsheh_vojta_prl_12}. Our numerical accuracy is not sufficient to determine whether the change of $\eta$ with $\Delta$ is a smooth crossover or a sharp transition somewhere around $\Delta=0.3$.

The anomalous exponent at the TLRO-to-QLRO transition at $T_{c1}$ also remains at its clean value for the weak disorder. For $\Delta \gtrapprox 0.3$, it rises slightly, but the increase is minuscule, smaller than the error bars. With the current numerical accuracy, we therefore cannot determine with certainty whether this exponent becomes disorder-dependent before the multicritical point is reached.

\begin{figure}%
    \centering
    \includegraphics[width=\columnwidth]{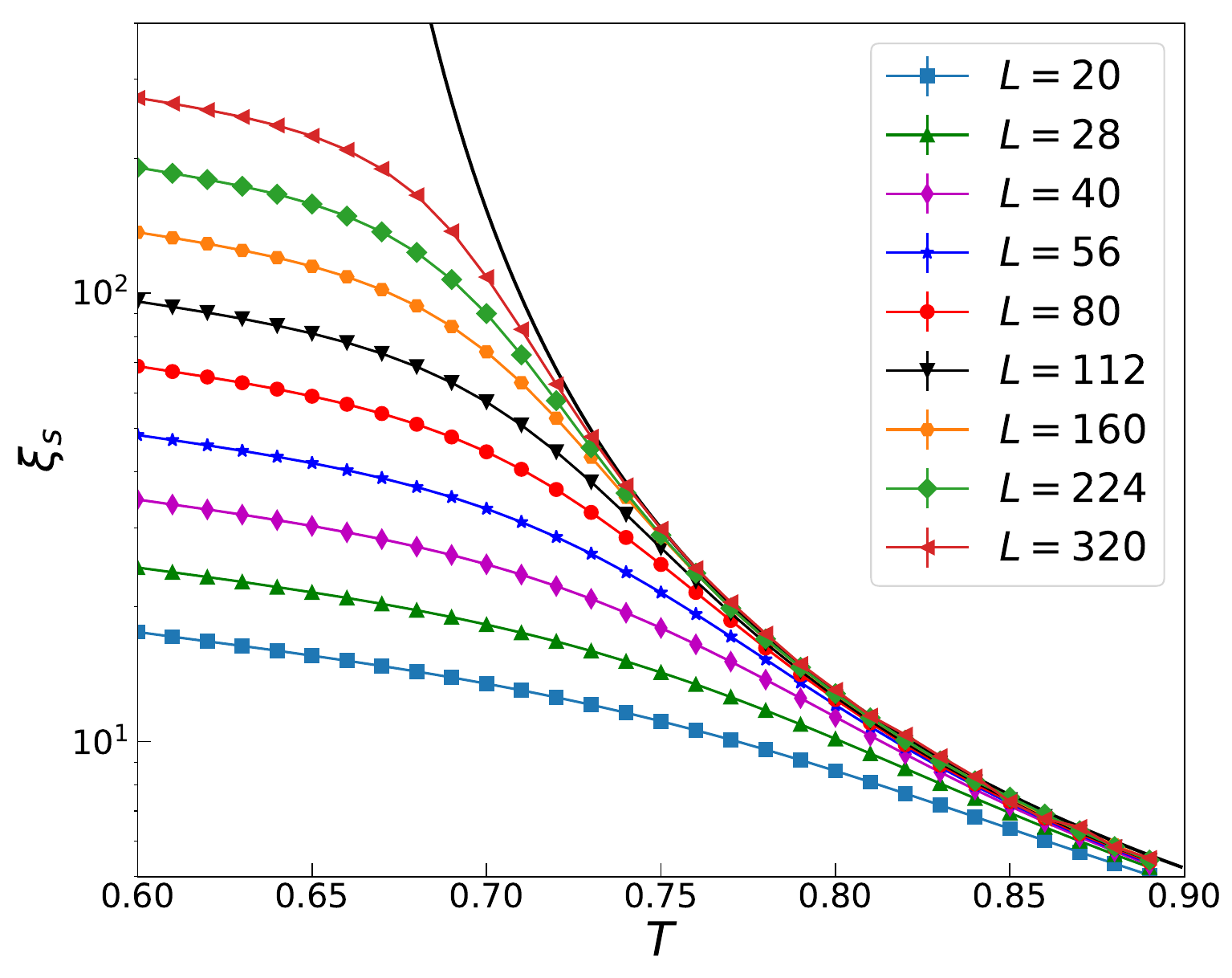}
    \caption{Spatial correlation length $\xi_s$ vs.\ temperature $T$ in the paramagnetic phase for $\Delta=0.3$. The solid line is a fit of the data for the largest system size with the BKT form (\ref{eq_corr_crit}) in the temperature range from 0.73 to 0.9. The value of $T_{c2}$ is fixed at 0.625, as found via the Binder cumulant crossings in Fig.\ \ref{fig:extrapolation_tcs}.}
    \label{fig:corrlen_critical}
\end{figure}

Finally, we confirm the BKT character of the paramagnetic-to-QLRO phase transition at $T_{c2}$ by studying the dependence of the correlation length $\xi_s$ on the distance from the transition point. Close to a BKT transition, $\xi_s$ is expected to follow the functional form
\begin{equation}
\label{eq_corr_crit}
    \xi_s = A e^{B(T-T_{c2})^{-1/2}}~.
\end{equation}
Here, $A$ and $B$ are non-universal constants. In Fig.~\ref{fig:corrlen_critical}, we plot the correlation length $\xi_s$ as a function of temperature in the paramagnetic phase for various system sizes at $\Delta = 0.3$.
The figure shows that $\xi_s$ follows the predicted exponential form until finite-size effects take over close to the transition point.

%%%%%%%%%%%%%%%%%%%%%%%%%%%%%%%%%%%%%%%%%%%%%%%%%%%%%%%%%%%%%%%%
\subsection{Critical behavior in the strong-disorder regime}
\label{subsec:4-d}
%%%%%%%%%%%%%%%%%%%%%%%%%%%%%%%%%%%%%%%%%%%%%%%%%%%%%%%%%%%55%%%

\begin{figure*}
 \includegraphics[width=\textwidth]{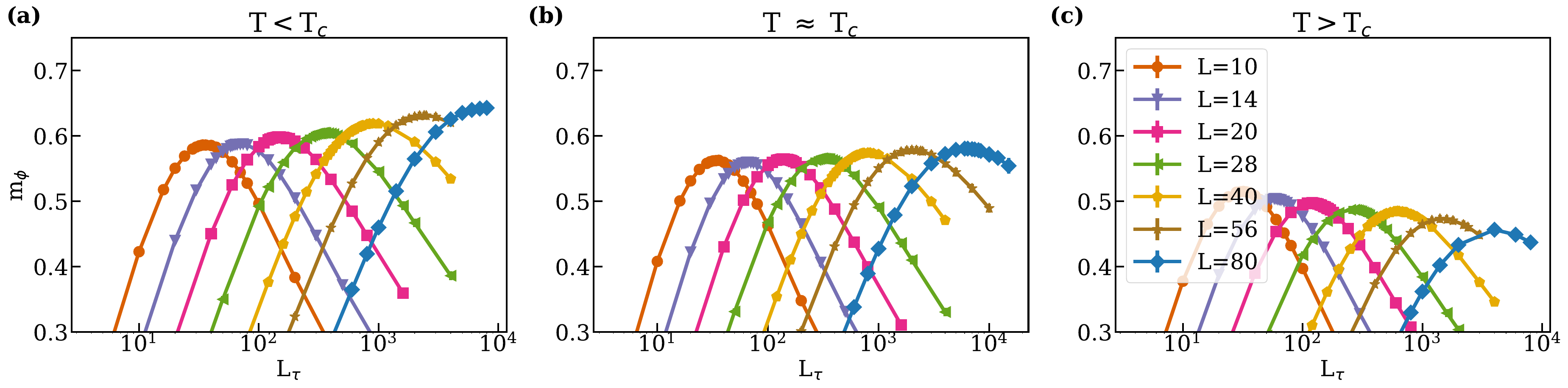}
        \caption{Clock order parameter $m_{\phi}$ as a function of the imaginary-time system size $L_\tau$ for different spatial system sizes $L$ and disorder strength $\Delta=1.5$. The temperatures are (a) $T=0.196<T_{c}$, (b) $T=0.200\approx T_{c}$ and (c) $T=0.208>T_{c}$. The peak values of $m_\phi$ are independent of $L$ at $T_{c}$ whereas they decrease with $L$ for $T>T_c$ and increase with $L$ for $T<T_{c}$. The statistical errors are of the order of the symbol size or smaller.}
    \label{fig:mphi_leveling_domes}
\end{figure*}

We now turn our attention to the strong-disorder regime, i.e., disorder strengths above the multicritical value $\Delta \approx 0.5$, where the system features a direct transition between the TLRO and paramagnetic phases. To explore this region, we have performed simulations for several disorder strengths, viz., $\Delta=0.7,1.0,1.5$ and $2.0$. Based on the SDRG of Senthil and Majumdar \cite{senthil_majumdar_prl_96}, the critical behavior at sufficiently large $\Delta$ is expected to be of the infinite-randomness type with activated rather than conventional power-law dynamical scaling. To test this prediction, we perform the anisotropic FSS analysis discussed in Sec.\ \ref{subsec:FSS}.
The results are shown in Fig.~\ref{fig:mphi_leveling_domes} for the prototypical case of $\Delta=1.5$.
To find the critical temperature $T_c$, we use the fact that the $m_{\phi}$ vs.\ $L_\tau$ curve (at fixed $L$ and $T$) develops a maximum at $L_\tau^{\rm max}$ which marks the ``optimal shape'' that keeps the second argument of the scaling functions in Eqs.\ (\ref{eqn:Binder_scal}) to (\ref{eqn:clock_op_scal_act}) constant. The value of $m_\phi$ at the maximum is independent of the system size $L$ right at criticality \cite{sknepnek_vojta_prl_04,rieger_young_prl_94}, as is the case in Fig.\ \ref{fig:mphi_leveling_domes}(b). In contrast, this maximum value increases with $L$ for $T<T_c$ as illustrated in Fig.\ \ref{fig:mphi_leveling_domes}(a), whereas it decreases with $L$ for $T>T_c$, see Fig.\ \ref{fig:mphi_leveling_domes}(c).
Following this methodology, we estimate the transition temperatures to be $0.380(6)$, $0.294(6)$, $0.200(8)$, and $0.140(10)$ for $\Delta=0.7$, $1.0$, $1.5$ and $2.0$, respectively.

\begin{figure}
   \includegraphics[width=\columnwidth]{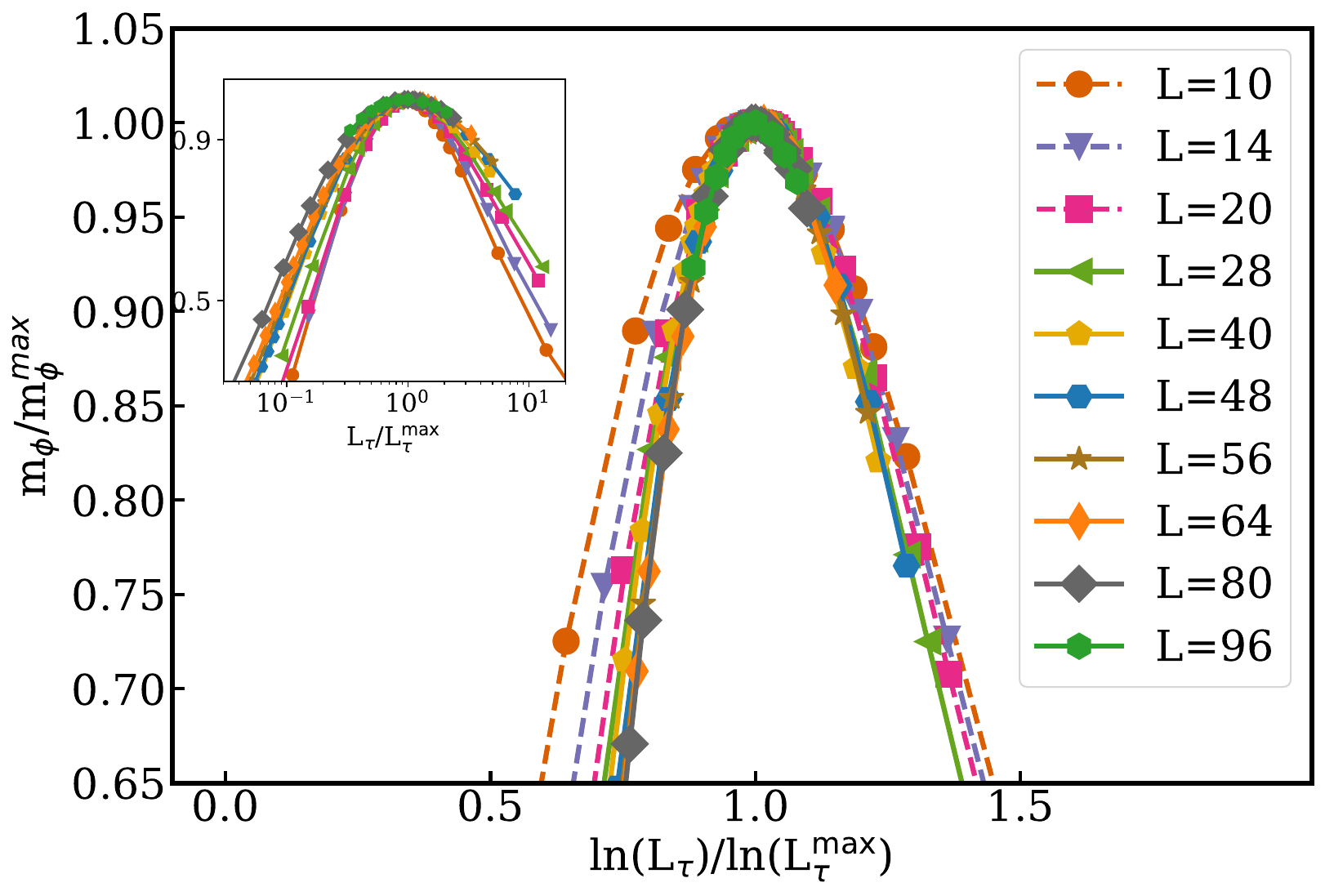}
       \caption{Scaling plot of $m_\phi/m_\phi^{\rm max}$ vs.\ $\ln(L_\tau)/\ln(L_\tau^{\rm max})$ at the critical temperature for $\Delta = 1.5$ according to the activated scaling ansatz (\ref{eqn:clock_op_scal_act}).  The dashed lines represent the data of the smallest system sizes which do not collapse perfectly due to the corrections to scaling. Inset: Same data plotted according to the power-law scaling ansatz (\ref{eqn:clock_op_scal}). The data do not collapse onto each other but broaden with increasing system size.}
    \label{fig:scaling_delta_1.5}
\end{figure}

To distinguish conventional power-law dynamical scaling from activated dynamical scaling, we investigate the scaling collapse of the $m_\phi$ vs.\ $L_\tau$ curves for $\Delta=1.5$ in Fig.\ \ref{fig:scaling_delta_1.5}.
The figure clearly shows that the data collapses very well when plotted according to the activated scaling ansatz (\ref{eqn:clock_op_scal_act}). Small deviations from perfect collapse for the smallest system sizes can be attributed to corrections to scaling stemming from the crossover towards infinite-randomness criticality. In contrast, the data do not collapse when plotted according to the conventional power-law scaling ansatz (\ref{eqn:clock_op_scal}) but rather broaden with increasing system size. We therefore conclude that the critical behavior is of infinite-randomness type. We have repeated this analysis for $\Delta$ = $1.0$, and $2.0$, with analogous results. The data collapse using the activated scaling scenario gets increasingly better, and extends to smaller system sizes as the disorder strength increases. This implies the presence of a cross-over length scale which decreases with increasing disorder strength.

\begin{figure}
 \includegraphics[width=\columnwidth]{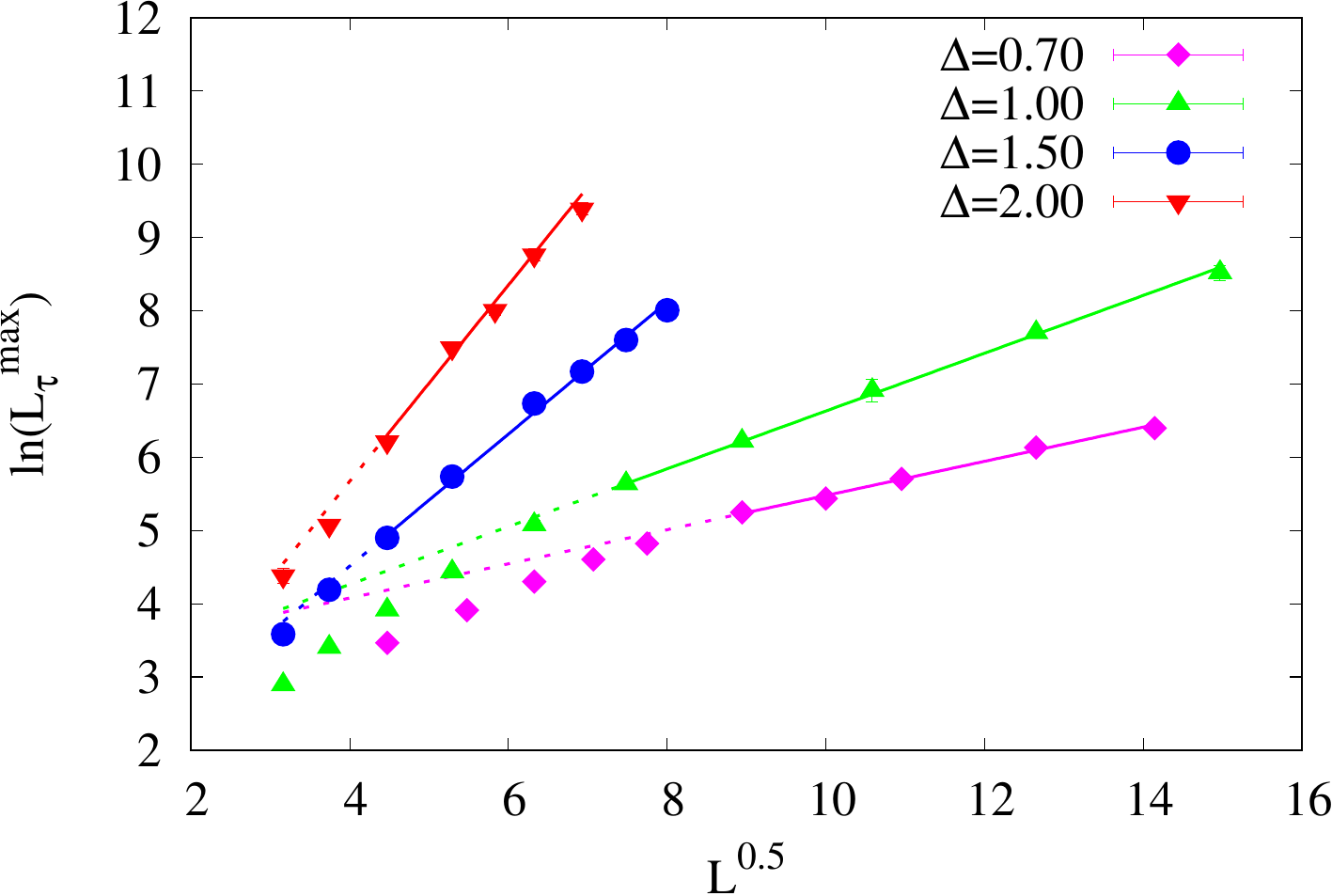}
    \caption{$\ln(L_{\tau}^{\rm max})$ vs.\ $L^{0.5}$ at criticality for several disorder strengths in the strong-disorder regime. The solid lines are fits with the activated scaling relation $\ln(L_{\tau}^{\rm max})=a + b L^{\psi}$. Here $\psi$ is fixed at the theoretical value of 0.5, and $a$ and $b$ are fit parameters. The dotted lines mark smaller system sizes that are not included in the fits.}
    \label{fig:fit_psi_activated}
\end{figure}

To compare the critical behavior quantitatively with the predictions of Senthil and Majumdar \cite{senthil_majumdar_prl_96},
we analyze the dependence of $L_\tau^{\rm max}$ on $L$ in Fig.~\ref{fig:fit_psi_activated}.
The figure shows that the data follow the predicted activated scaling relation $\log(L_{\tau}^{\rm max}) \sim L^{\psi}$ with $\psi=1/2$ for all disorder strengths we studied in the strong-disorder regime, provided the system size is larger than a disorder-dependent crossover length scale.

Specifically, the data for the strongest disorders,  $\Delta =2.0$ and 1.5, follow the infinite-randomness prediction for all $L\ge 20$. The $\Delta=1$ data follow the prediction for $L \ge 56$, whereas the data for the weakest disorder, $\Delta =0.7$, only follow the prediction if $L \ge 80$. This shows that the cross-over length scale increases with decreasing disorder (as the multicritical point is approached).

\begin{figure}
  \includegraphics[width=\columnwidth]{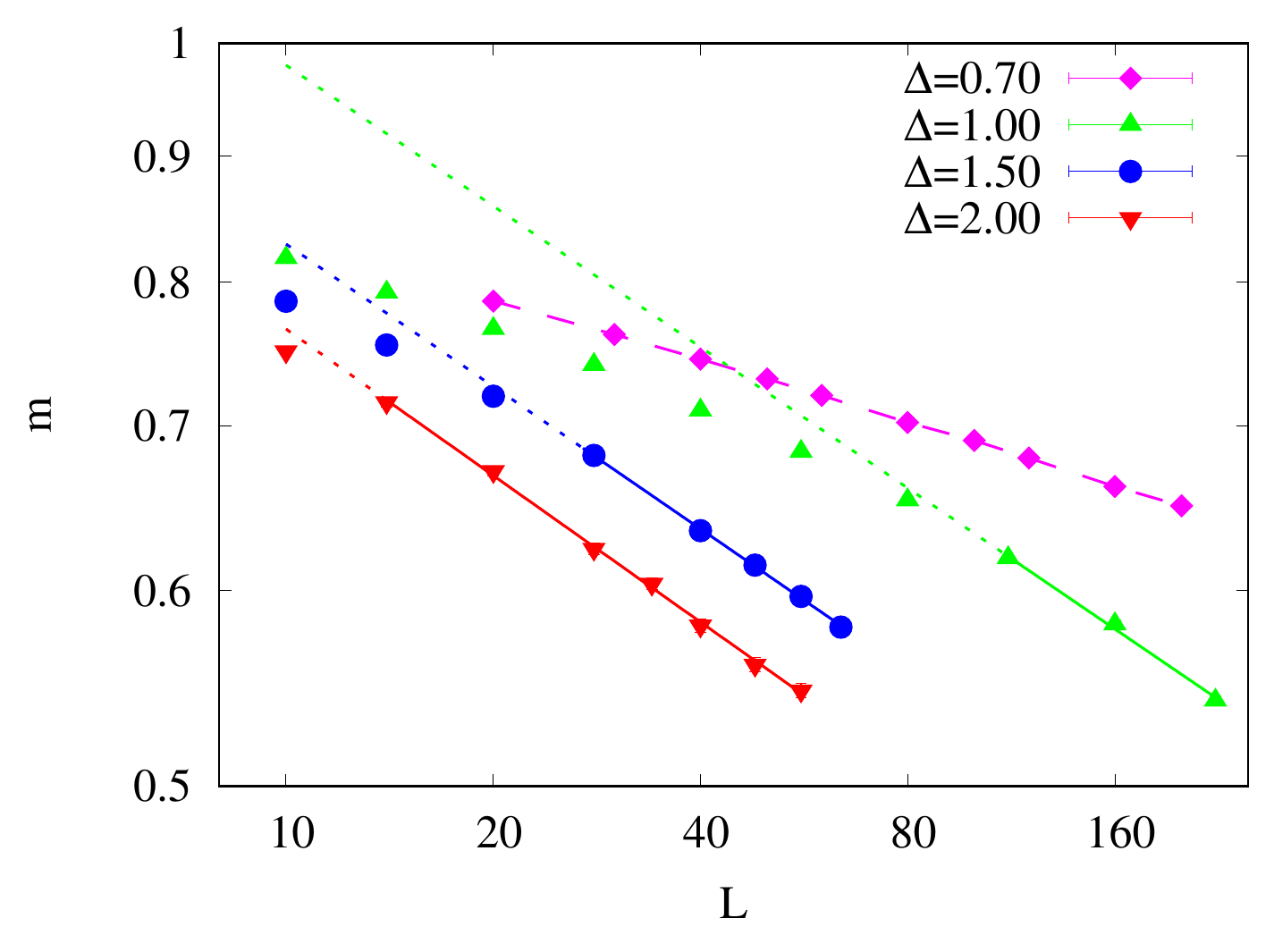}
    \caption{ Average magnetization $m$ at criticality vs.\ system size $L$ for several disorder strengths $\Delta$ in the strong disorder regime. Solid lines are fits with  $m= a L^{-\frac{\beta}{\nu}}$ with $\beta/\nu$
    and $a$ fit parameters. The fits yield $\beta/\nu=0.19(1)$ for disorder strengths $\Delta \ge 1.0$. The dotted lines mark smaller system sizes that are not included in the fits. The data for $\Delta=0.7$ (dashed line) do not appear to reach the asymptotic behavior for the available system sizes. Thus, we do not show a power-law fit of these data }
    \label{fig:fitbetabynu_activated}
\end{figure}

Finally, we analyze the system-size dependence of the average magnetization at criticality (using the optimal shapes, i.e., $L_\tau = L_\tau^{\rm max}$). Figure \ref{fig:fitbetabynu_activated} shows that
the data for disorder strengths $\Delta \ge 1.0$ follow the SDRG prediction $m \sim L^{-\beta/\nu}$ with critical exponent $\beta/\nu=0.19$ for system sizes above a crossover scale that increases rapidly with decreasing disorder.
The data for the weakest disorder we study in this regime, $\Delta=0.7$, show the beginning of a crossover (manifest in the small downward curvature of the curve with increasing $L$), but the crossover length appears to be larger than the maximum system size in our simulations

This implies that a power-law fit of the data for $\Delta =0.7$ would only yield an effective ($L$-dependent) exponent value rather than the true asymptotic exponent
(analogous to a fit of the $\Delta=1.00$ data for $L < 96$).

We thus conclude that the magnetization data for $\Delta =0.7$ are compatible with the infinite-randomness scenario predicted by the SDRG, but we cannot exclude a different asymptotic behavior.

Within the infinite-randomness scenario, the critical point is expected to be accompanied by quantum Griffiths phases \cite{fisher_prl_92,*fisher_prl_95,RiegerYoung96,GuoBhattHuse96, hoyos_kotabage_prl_07,*vojta_kotabage_prb_09} that feature non-universal power-law singularities of observables not just at criticality but in an entire region around the transition.
In order to identify quantum Griffiths behavior in our simulations, we follow the method employed in Ref.\ \cite{HrahshehBarghathiVojta11} and study the susceptibility $\chi$ as a function of $L_\tau$ for large fixed $L$ close to (but not exactly at) the transition. For the original quantum Hamiltonian (\ref{eqn:quantum-hamiltonian}), this corresponds to studying the temperature dependence of $\chi$ slightly off criticality. Specifically, we compute $\chi$ for disorder strength $\Delta=2.0$,
spatial system size $L=800$, and temporal sizes $L_\tau=10$ to $80$ in the temperature region from $T=0.125$ to $0.25$. For a range of temperatures close to $T_c=0.14$ (as found from the scaling analysis of $m_\phi$ above), the susceptibility follows the expected non-universal power law $\chi \sim L_\tau^{1\pm 1/z'}$ where $z'$ is the Griffiths dynamical exponent. Within the infinite-randomness scenario, $z'$ is expected to diverge at $T_c$, leading to a linear relation between $\chi$ and $L_\tau$.
Our simulation data do not quite agree with this, they give $\chi \sim L_\tau^{1.2}$ at $T=0.14$.
We attribute this deviation to finite-size effects due to the relatively small $L_\tau$ values from 10 to 80 \footnote{In the Griffiths phase, the susceptibility contains a regular bulk contribution and the singular rare region contribution that produces the Griffiths singularities. For small $L_\tau$, corrections to the Griffiths physics stemming from the regular part are still sizable.}.

It is interesting to compare our strong evidence for a Griffiths phase in the form of non-universal power-law behavior of the susceptibility with the results of Ref.~\cite{Chatelain} for the $(2+1)$ dimensional disordered clock model. In Ref.\ \cite{Chatelain}, the evidence for the very existence of the Griffiths phase is inconclusive because the data do not consistently show non-universal power-law behavior.  This may stem from the fact that the strengths of Griffiths singularities decreases with increasing dimensionality \cite{THILL1995321,GuoBhattHuse96} implying that stronger disorder and/or larger sizes may be necessary to see the Griffiths phase in $(2+1)$ dimensions.

%%%%%%%%%%%%%%%%%%%%%%%%%%%%%%%%%%%%%%%%%%%%%%%%%%%%%%%%%%%%%%%%%%%%
\section{Conclusion and Summary}
\label{sec-5}
%%%%%%%%%%%%%%%%%%%%%%%%%%%%%%%%%%%%%%%%%%%%%%%%%%%%%%%%%%%%%%%%%%%%

In this manuscript, we have investigated the impact of quenched random disorder on the phases and phase transitions of the one-dimensional quantum q-state clock model, focusing on $q=6$ which serves as a test bed for all $q>4$ cases.
To this end, we have performed large-scale Monte Carlo simulations of a classical clock model with a perfectly correlated (columnar) disorder that arises from the quantum-to-classical mapping of the quantum model's partition function. Our results have established that the disorder is inimical to the emergent QLRO phase that separates the TLRO phase from the paramagnetic phase in the clean system. Specifically, the intermediate QLRO phase shrinks with increasing disorder strength,  finally resulting in a multicritical point beyond which there is a direct phase transition from the clock-ordered (TLRO) phase to the paramagnetic phase.

The bulk of our results were obtained using the power-law disorder distribution (\ref{eq_powerlaw_disorder}). This distribution becomes arbitrarily broad (even on a logarithmic scale) for $\Delta\to \infty$,  We have tested the influence of the disorder distribution by performing simulations with a binary distribution as implemented in Ref.\ \cite{hrahsheh_vojta_prl_12}. In agreement with the expected universality of the phase diagram, these simulations show the same shrinking of the QLRO phase with increasing disorder strengths. However, the crossover to the strong disorder regime is slower, and the asymptotic behavior is not reached for the numerically accessible system sizes. We expect analogous behavior for other ``less broad'' distributions such as the box distribution.

We have also characterized the critical behavior along the different phase boundaries in the disorder-(classical) temperature phase diagram. In agreement with the Harris criterion \cite{harris_jpcssp_74}, the critical behaviors of the BKT transitions from the TLRO phase to the QLRO phase as well as from the QLRO phase to the paramagnetic phase are stable against weak disorder. As the disorder strength is increased, both transitions retain the BKT character with a dynamical exponent $z=1$ all the way to the multicritical point (or, at least, close to it). However, the anomalous exponent $\eta$ of the QLRO-to-paramagnet transition deviates from its clean value and becomes disorder dependent, analogous to the behavior observed in a disordered (1+1)-dimensional XY model \cite{hrahsheh_vojta_prl_12}. For the TLRO-to-QLRO transition, in contrast, the changes of $\eta$, if any, are small and within the error bars of our simulations.

For disorder strengths well above the multi-critical point, our results have demonstrated that the critical behavior of the direct transition between the TLRO and paramagnetic phases is of exotic infinite-randomness type and falls into the random transverse-field Ising universality class, as predicted by the SDRG analysis of Ref.~\cite{senthil_majumdar_prl_96}. Our results have also provided numerical evidence for a quantum Griffiths phase associated with the infinite-randomness critical point.

The critical behavior of the multi-critical point itself is difficult to determine numerically because the crossovers between the three different critical behaviors on the adjacent phase boundaries lead to a strong and complex finite-size correction. Resolving these with a Monte Carlo approach would require significantly larger system sizes than we were able to access in our simulations. The same also applies to the interesting question of whether the infinite-randomness critical behavior takes over right after the disorder strength is increased beyond its multicritical value. While our numerical data show no indication of an intermediate regime between the multicritical point and infinite-randomness criticality at the largest disorders, a complete analysis of the vicinity of the multicritical point remains a task for the future.

Interesting questions arise for higher-dimensional quantum clock models and their classical counterparts. The two-dimensional $q$-state quantum clock model and its three-dimensional classical counterpart do not have an intermediate phase but a direct transition from the clock-ordered phase to the paramagnetic phase. For $q \ge 4$, the critical point has an emerging XY symmetry. However, this symmetry is broken down to a discrete $Z_q$ symmetry in the ordered phase by means of a dangerously irrelevant variable \cite{blankschtein_ma_prb_84,Oshikawa00,LouSandvikBalents07}. What happens to this scenario in the presence of quenched disorder? A recent study employing a numerical SDRG calculation concluded that the critical behavior is of the IRFP type \cite{Chatelain}. As discussed in Sec.\ \ref{subsec:4-d}, the evidence for quantum Griffith behavior in the vicinity of the transition was inconclusive. An investigation along the lines of our study may help address this problem and potentially unravel surprising disorder-induced cross-over effects.

\begin{acknowledgments}
The simulations were performed on the Pegasus and Foundry clusters at Missouri S\&T and the Aqua cluster at IIT Madras. We thank Ambuj Jain, Pranay Patil, Shashikant Singh, and Jos\'{e} A. Hoyos for the helpful discussions.  P.K.V.\ acknowledges support via the IIE Travel award by the Global Engagement Office, IIT Madras. R.N.\ and P.K.V.\ also acknowledge funding from the Center for Quantum Information Theory in Matter and Spacetime, IIT Madras and from the Department of Science and Technology, Govt. of India, under Grant No. DST/ICPS/QuST/Theme-3/2019/Q69, as well as support from the Mphasis F1 Foundation via the Centre for Quantum Information, Communication, and Computing (CQuICC). G.K. and T.V. acknowledge support from the National Science Foundation under grant nos. DMR-1506152, DMR-1828489, OAC-1919789. T.V. also acknowledges support for a visit at IIT Madras, where part of the work was completed, via a Visiting Faculty Fellowship of the Office of Global Engagement IIT Madras.
\end{acknowledgments}

%%%%%%%%%%%%%%%%%%%%%%%%%%%%%%%%%%%%%%%%%%%%%%%%%%%%%%%%%%%%%%%%%%%%%%%%
\appendix
\section{Quantum-to-Classical Mapping}
%%%%%%%%%%%%%%%%%%%%%%%%%%%%%%%%%%%%%%%%%%%%%%%%%%%%%%%%%%%%%%%%%%%%%%%%%%%%

In this section, we discuss the quantum-to-classical mapping that relates the partition function of a $d$-dimensional quantum system to that of a classical system in
$D=d+1$ dimensions. It is based on expressing the quantum partition function as a path integral in $(d+1)$-dimensional Euclidean space-time. This is formally always possible, e.g., via a Trotter decomposition of the partition function. However, the question of whether or not the action of this path integral can be interpreted as a classical Hamiltonian in $(d+1)$ space dimensions depends on the problem at hand. In some cases (including ours), the action is real and leads to positive statistical weights. It can thus be understood as a classical Hamiltonian. In other cases, such as the bosonic Hubbard model at non-integer filling, the action leads to negative or complex weights and cannot be interpreted as the Hamiltonian of a classical system \cite{Wallin, Fishers_dirty_boson}.

We now derive the mapping from the one-dimensional quantum clock Hamiltonian (\ref{eqn:quantum-hamiltonian}) to the equivalent two-dimensional classical clock model (\ref{eq_Hamiltonian_cl}). For simplicity, let us consider the translational invariant case with $J_{l}=J$ and $h_{l}=h$.
We rewrite the Hamiltonian (\ref{eqn:quantum-hamiltonian}) in terms of generalized Pauli matrices as follows,
\begin{equation}
\begin{split}
    H&=-\frac J 2 \sum_{i}^{N}\left[\hat{\tau}_{i}\hat{\tau}^{\dagger}_{i+1}+\hat{\tau}^{\dagger}_{i}\hat{\tau}_{i+1}\right]- \frac h 2 \sum_{i}^{N}\left[\hat{\Gamma}_{i}+\hat{\Gamma}^{\dagger}_{i}\right]\\ &=
    H_{0}+H_{1}~.
\end{split}
\end{equation}
The matrix forms of the clock operators $\hat{\tau}$ and the shift operators $\hat{\Gamma}$ in the clock state basis $\ket{p}$ with $p=0, \ldots, q-1$ are given by
\begin{equation}
 \hat{\tau}=\left(\begin{array}{ccccc}
1 & 0 & 0 & \cdots & 0 \\
0 & \omega & 0 & \cdots & 0 \\
0 & 0 & \omega^2 & \cdots & 0 \\
\vdots & \vdots & \vdots & \ddots & \vdots \\
0 & 0 & 0 & \cdots & \omega^{(q-1)}
\end{array}\right)
\end{equation}
and
\begin{equation}
\hat{\Gamma}=\left(\begin{array}{ccccc}
0 & 1 & 0 & \cdots & 0 \\
0 & 0 & 1 & \cdots & 0 \\
0 & 0 & \ddots & 1 & 0 \\
\vdots & \vdots & \vdots & \ddots & \vdots \\
1 & 0 & 0 & \cdots & 0
\end{array}\right) ~,
\end{equation}
where $\omega=e^{2\pi  i/q}$. In other words,
\begin{eqnarray}
    \hat{\tau}\ket{p} &=& \omega^{p}\ket{p}~, \\
    \hat{\Gamma}\ket{p} &= &\ket{(p-1)\mod q}~.
\end{eqnarray}
The clock and shift operators at a given lattice site satisfy the relations
\begin{equation}
    \hat{\tau}^{q}=I=\hat{\Gamma}^{q}\quad\text{ and}\quad \hat{\tau}\hat{\Gamma}=\omega\hat{\Gamma}\hat{\tau}~.
\end{equation}
The clock and shift operators commute at different sites.

Consider the partition function of the Hamiltonian (\ref{eqn:quantum-hamiltonian}) at a (quantum) temperature $T={1}/{\beta}$
\footnote{In this appendix, the inverse physical temperature of the quantum Hamiltonian (\ref{eqn:quantum-hamiltonian}) is denoted by $\beta$ while that of the mapped classical system (\ref{eq_Hamiltonian_cl}) is called $\beta_{cl}$. },
\begin{equation}
Z=\operatorname{Tr}\left(e^{-\beta H}\right)~.
\end{equation}
The operator $e^{-\beta H}$ can be regarded as the imaginary time evolution operator from imaginary time 0 to $\beta$. Using the standard Trotter decomposition technique, we can decompose $Z$ as
\begin{equation}
Z=\operatorname{Tr}(\underbrace{e^{-\Delta\tau H} e^{-\Delta\tau H} \ldots e^{-\Delta\tau H}}_{L_{\tau} \text { times }}) \quad\left(\text{  where }\Delta\tau=\frac{\beta}{L_{\tau}}\right)~.
\end{equation}
Here, $L_{\tau}$ is the number of time slices.
Now, we use the completeness condition
\begin{equation}
    I=\sum_{\{p=0,1,...q-1\}}\ket{p}\bra{p}
\end{equation}
with $\ket{p}=\ket{p_{1}}\bigotimes\ket{p_{2}}\bigotimes...\ket{p_{N}}$.
Imposing periodic boundary conditions in the imaginary time direction $\ket{p^{0}}=\ket{p^{L_{\tau}}}$, the partition function can be expressed as
\begin{equation}
\begin{aligned}
Z= & \sum_{\{p^{\tau}\}}\left\langle p^{0}\left|e^{-\Delta\tau H}\right| p^{L_{\tau}-1}\right\rangle\left\langle p^{L_{\tau}-1}\left|e^{-\Delta\tau H}\right| p^{L_{\tau}-2}\right\rangle \\
& \cdots\left\langle p^{\tau+1}\left|e^{-\Delta\tau H}\right| p^{\tau}\right\rangle \cdots\left\langle p^{1}\left|e^{-\Delta\tau H}\right| p^{0}\right\rangle
\end{aligned}~,
\end{equation}
where we have introduced, at each time step $\tau$, a complete set of states $\ket{p^\tau}$. We assume that $L_{\tau}$ is large so that $\Delta\tau$ is small and further decompose
\begin{equation}
e^{-\Delta\tau H}=e^{-\Delta\tau\left(H_0+H_1\right)} \simeq e^{-\Delta\tau H_0} e^{-\Delta\tau H_1}~.
\end{equation}

Now, consider one such matrix element :
\begin{equation}
\begin{aligned}
\left\langle p^{\tau+1}\left|e^{-\Delta\tau H}\right| p^{\tau}\right\rangle & \simeq\left\langle p^{\tau+1}\left|e^{-\Delta\tau H_0} e^{-\Delta\tau H_1}\right| p^{\tau}\right\rangle ~.
\end{aligned}
\end{equation}
We note that $\ket{p}$ is an eigenstate of $\hat{\tau}$ operator. Therefore
\begin{equation}
\begin{split}
    &e^{\hat{\tau}_{i}\hat{\tau}_{i+1}^{\dagger}}\ket{p_{i}p_{i+1}}=e^{\omega^{p_{i}}\omega^{-p_{i+1}}}\ket{p_{i}p_{i+1}}\quad \text{and}\\
    &e^{\hat{\tau}_{i}^{\dagger}\hat{\tau}_{i+1}}\ket{p_{i}p_{i+1}}=e^{\omega^{-p_{i}}\omega^{p_{i+1}}}\ket{p_{i}p_{i+1}}~.
\end{split}
\end{equation}
We have
\begin{equation}
\begin{split}
    &\omega^{p_{i}}\omega^{-p_{i+1}}=e^{\frac{2\pi i}{q}(p_{i}-p_{i+1})} \quad \text{and}\\
    &\omega^{-p_{i}}\omega^{p_{i+1}}=e^{-\frac{2\pi i}{q}(p_{i}-p_{i+1})}~.
    \end{split}
\end{equation}
From the above expressions, we can write
\begin{equation}
    e^{\frac{2\pi i}{q}(p_{i}-p_{i+1})}+e^{-\frac{2\pi i}{q}(p_{i}-p_{i+1})}=2\cos{\left[\frac{2\pi}{q}(p_{i}-p_{i+1})\right]}~.
\end{equation}
Hence, the matrix element can be written as
\begin{equation}
\begin{split}
    &\left\langle p^{\tau+1}\left|e^{-\Delta\tau H_0} e^{-\Delta\tau H_1}\right| p^{\tau}\right\rangle \\&=\left\langle p^{\tau+1}\left|e^{+J\Delta\tau \sum_{i}^{N}\cos{[\frac{2\pi}{q}(p_{i}^{\tau+1}-p_{i+1}^{\tau+1})]}} e^{-\Delta\tau H_1}\right| p^{\tau}\right\rangle~.
\end{split}
\end{equation}
Now, we act the remaining operator on the eigenstates. Consider
\begin{equation}
    \begin{split}
        &\left\langle p^{\tau+1}\left|e^{-\Delta\tau H_1}\right| p^{\tau}\right\rangle\\
        &=\left\langle p^{\tau+1}\left|e^{+\frac h 2 \Delta\tau\sum_{i}(\Gamma_{i}+\Gamma_{i}^{\dagger})}\right| p^{\tau}\right\rangle~.
    \end{split}
\end{equation}
Let us assume $\theta=\frac h 2 \Delta\tau$. Then,
\begin{equation}
e^{\theta\Gamma}=\sum_{n}\frac{(\theta\Gamma)^{n}}{n!}~.
\end{equation}
Here, we split the expansion into q different series as follows
\begin{equation}
    e^{\theta\Gamma}=\sum_{p}\frac{(\theta\Gamma)^{qp}}{(qp)!}+\sum_{p}\frac{(\theta\Gamma)^{qp+1}}{(qp+1)!}+...+\sum_{p}\frac{(\theta\Gamma)^{qp+q-1}}{(qp+q-1)!}
\end{equation}
Using the identity $\Gamma^{q}$=I, the expansion simplifies to
\begin{equation}
    e^{\theta\Gamma}=\sum_{p}\frac{\theta^{qp}}{(qp)!}+\sum_{p}\frac{\theta^{qp+1}\Gamma}{(qp+1)!}+...+\sum_{p}\frac{\theta^{qp+q-1}\Gamma^{q-1}}{(qp+q-1)!}~`
\end{equation}
which we represent as
\begin{equation}
    e^{\theta\Gamma}=f_{0}(\theta)+f_{1}(\theta)\Gamma+...+f_{q-1}(\theta)\Gamma^{q-1}~,
\end{equation}
where we denote
\begin{equation}
    f_{p-1}(\theta)=\Lambda e^{\gamma\omega^{p-1}}~.
\end{equation}
Now, let's evaluate the following elements,
\begin{equation}
    \begin{split}
        &\left\langle p^{\tau+1}\left|e^{\theta\Gamma}\right| p^{\tau}\right\rangle\\
        &=\left\langle p^{\tau+1}\left|f_{0}(\theta)I+f_{1}(\theta)\Gamma+...+f_{q-1}(\theta)\Gamma^{q-1}\right| p^{\tau}\right\rangle\\
        &= \begin{cases}f_{0}(\theta) & \text { if } p^{\tau+1} = p^{\tau} \\ f_{1}({\theta}) & \text { if } p^{\tau+1} = (p^{\tau}+1) \\ ... \\
        f_{q-1}(\theta) & \text{ if } p^{\tau+1}=(p^{\tau}+q-1)\end{cases}\\
        &= \begin{cases}\Lambda e^{\gamma\omega^{0}} & \text { if } p^{\tau+1} = p^{\tau} \\ \Lambda e^{\gamma\omega^{1}} & \text { if } p^{\tau+1} = (p^{\tau}+1) \\ ... \\
        \Lambda e^{\gamma\omega^{q-1}} & \text{ if } p^{\tau+1}=(p^{\tau}+q-1)\end{cases}\\
        &=\Lambda e^{\gamma\omega^{-(p^{\tau}-p^{\tau+1})}}~.
    \end{split}
\end{equation}
Similarly,
\begin{equation}
        \left\langle p^{\tau+1}\left|e^{\theta\Gamma^{\dagger}}\right| p^{\tau}\right\rangle= \Lambda e^{\gamma\omega^{(p^{\tau}-p^{\tau+1})}}~.
\end{equation}
Therefore,
\begin{equation}
\begin{split}
    \left\langle p^{\tau+1}\left|e^{\theta(\Gamma+\Gamma^{\dagger})}\right| p^{\tau}\right\rangle \\&= \Lambda^{2} e^{\gamma\omega^{(p^{\tau}-p^{\tau+1})}}e^{\gamma\omega^{-(p^{\tau}-p^{\tau+1})}} \\
    &=\Lambda^{2} e^{\gamma \left[e^{\frac{2\pi i}{q}{(p^{\tau}-p^{\tau+1})}}+e^{-\frac{2\pi i}{q}{(p^{\tau}-p^{\tau+1})}}\right]}\\
    &=\Lambda^{2} e^{2\gamma \left[\cos{\frac{2\pi}{q}{(p^{\tau}-p^{\tau+1})}}\right]}~,
\end{split}
\end{equation}
which implies that,
\begin{equation}
    \left\langle p^{\tau+1}\left|e^{\frac h 2 \Delta\tau\sum_{i}(\Gamma_{i}+\Gamma_{i}^{\dagger})}\right| p^{\tau}\right\rangle=\Lambda^{2N} e^{2\gamma \sum_{i}^{N}\left[\cos{\frac{2\pi}{q}{(p^{\tau}_{i}-p^{\tau+1}_{i})}}\right]}~.
\end{equation}
The partition function can be rewritten to the following form
\begin{equation}
\begin{split}
    Z= & \Lambda^{2NL_{\tau}} e^{\sum_{i}^{N}\sum_{\tau}^{L_{\tau}}[J\Delta\tau \cos{[\frac{2\pi}{q}(p_{i}^{\tau}-p_{i+1}^{\tau})]}+2\gamma \cos{[\frac{2\pi}{q}{(p^{\tau}_{i}-p^{\tau+1}_{i})]}}]}~.
\end{split}
\end{equation}
This is equivalent to the partition function of a two-dimensional classical clock model with the following Hamiltonian:
\begin{equation}
\begin{split}
    H_{\rm cl}=&-\sum_{i,\tau} J^{s}\cos\left[\frac{2\pi(p_{i,\tau}-p_{i+1,\tau})}{q}\right]\\
    &-\sum_{i,\tau}J^{t}\cos\left[\frac{2\pi(p_{i,\tau}-p_{i,\tau+1})}{q}\right]~.
\end{split}
\end{equation}
The couplings are connected via the relation
\begin{equation}
    \beta_{\rm cl}J_{s}=J\Delta\tau, \quad \beta_{\rm cl}J_{t}=2\gamma ~.
\end{equation}
We have
\begin{equation}
    \begin{split}
        &e^{\frac h 2 \Delta\tau}=\Lambda\left[\sum_{p=0}^{q-1}e^{\gamma\omega^{p}}\right] \quad \text{and}\\
        &e^{-\frac h 2 \Delta\tau}=\Lambda\left[\sum_{p=0}^{q-1}(-1)^{p}e^{\gamma\omega^{p}}\right]~.
    \end{split}
\end{equation}
Therefore the relation between $h$ and $\gamma$ can be obtained as,
\begin{equation}
    e^{-h\Delta\tau}=\frac{\sum_{p=0}^{q-1}(-1)^{p}e^{\gamma\omega^{p}}}{\sum_{p=0}^{q-1}e^{\gamma\omega^{p}}}~.
\end{equation}
For the case of the Ising model ($q=2$), this reduces to
\begin{equation}
    e^{-h\Delta\tau}=\frac{e^{\gamma}-e^{-\gamma}}{e^{\gamma}+e^{-\gamma}}=\tanh{\gamma}~.
\end{equation}
Note that temperature is not the tuning parameter of the quantum phase transition in the quantum Hamiltonian. However, changing the (inverse) classical temperature $\beta_{\rm cl}$ in the mapped classical model corresponds to
changing $h/J$ in the quantum Hamiltonian. Thus, universal properties of the quantum phase transitions in the quantum clock model can be obtained from the finite temperature transitions of the corresponding 2D classical clock model.

\bibliography{references.bib}   % name your BibTeX data base
\end{document}